\documentclass[12pt,preprint]{aastex}
\usepackage{graphicx}

\newcommand{\h}{$h^{-1}$}
\newcommand{\um}{$\mu$m}

\newcommand{\spitzer}{\emph{Spitzer}}
\newcommand{\iras}{\emph{IRAS}}
\newcommand{\iso}{\emph{ISO}}
\newcommand{\cobe}{\emph{COBE}}

\begin{document}

\def\sarc{$^{\prime\prime}\!\!.$}
\def\arcsec{$^{\prime\prime}$}
\def\arcmin{$^{\prime}$}
\def\degr{$^{\circ}$}
\def\seco{$^{\rm s}\!\!.$}
\def\ls{\lower 2pt \hbox{$\;\scriptscriptstyle \buildrel<\over\sim\;$}} 
\def\gs{\lower 2pt \hbox{$\;\scriptscriptstyle \buildrel>\over\sim\;$}} 
\def\spose#1{\hbox to 0pt{#1\hss}}
\def\simlt{\mathrel{\spose{\lower 3pt\hbox{$\mathchar"218$}}
     \raise 2.0pt\hbox{$\mathchar"13C$}}}
\def\simgt{\mathrel{\spose{\lower 3pt\hbox{$\mathchar"218$}}
     \raise 2.0pt\hbox{$\mathchar"13E$}}}
 
\title{Mid-Infrared Galaxy Luminosity Functions from the AGN and Galaxy Evolution Survey}

\author{X.~Dai\altaffilmark{1,8}, R.~J.~Assef\altaffilmark{1}, C.~S.~Kochanek\altaffilmark{1}, M.~Brodwin\altaffilmark{2}, M.~J.~I.~Brown\altaffilmark{3}, N.~Caldwell\altaffilmark{4}, R.~J.~Cool\altaffilmark{5}, A.~Dey\altaffilmark{2}, P.~Eisenhardt\altaffilmark{6}, D.~Eisenstein\altaffilmark{5}, A.~H.~Gonzalez\altaffilmark{7}, B.~T.~Jannuzi\altaffilmark{2}, C.~Jones\altaffilmark{4}, S.~S.~Murray\altaffilmark{4} and D.~Stern\altaffilmark{6}}

\altaffiltext{1}{\small Department of Astronomy, The Ohio State University, Columbus, OH 43210, USA}
\altaffiltext{2}{\small NOAO, 950 N. Cherry Ave., P.O. Box 26732, Tucson, AZ 85726, USA}
\altaffiltext{3}{\small School of Physics, Monash University, Clayton 3800, Victoria, Australia}
\altaffiltext{4}{\small Harvard/Smithsonian Center for Astrophysics, 60 Garden St., MS-67, Cambridge, MA 02138, USA}
\altaffiltext{5}{\small Steward Observatory, University of Arizona, 933 N Cherry Ave., Tucson, AZ 85121, USA}
\altaffiltext{6}{\small Jet Propulsion Laboratory, California Institute of Technology, Pasadena, CA 91109, USA}
\altaffiltext{7}{\small Department of Astronomy, University of Florida, Gainesville, FL 32611-2055, USA}
\altaffiltext{8}{\small Department of Astronomy, University of Michigan, Ann Arbor, MI, 48109, USA}
\email{xdai@umich.edu}

\begin{abstract}
We present galaxy luminosity functions at 3.6, 4.5, 5.8, and 8.0\um\ measured by
combining photometry from the IRAC Shallow Survey with redshifts
from the AGN and Galaxy Evolution Survey 
of the NOAO Deep Wide-Field Survey Bo\"otes field. 
The well-defined IRAC samples contain 3800--5800 galaxies for the 3.6--8.0\um\ bands
with spectroscopic redshifts and $z < 0.6$. 
We obtained relatively complete luminosity functions in the local redshift bin of $z < 0.2$ 
for all four IRAC channels that are well fit by Schechter functions.
After analyzing the samples for the whole redshift range, we found significant evolution in the 
luminosity functions for all four IRAC channels that can be fit as an evolution in $M_*$ with redshift,
$\Delta M_* = Qz$. 
While we measured $Q=1.2\pm0.4$ and $1.1\pm0.4$ in the 3.6 and 4.5\um\ bands consistent with the predictions
from a passively evolving population, we obtained $Q=1.8\pm1.1$ in the 8.0\um\ band consistent
with other evolving star formation rate estimates. 
We compared our LFs with the predictions of semi-analytical galaxy formation 
and found the best agreement at 3.6 and 4.5\um, 
 rough agreement at 8.0\um, and a large mismatch at 5.8\um.
These models also predicted a comparable $Q$ value to our luminosity functions at 8.0\um, but
predicted smaller values at 3.6 and 4.5\um.
We also measured the luminosity functions separately for early and late-type galaxies. 
While the luminosity functions of late-type galaxies resemble those for the total population, 
the luminosity functions of early-type galaxies in the 3.6 and 4.5\um\ bands indicate deviations from 
the passive evolution model, especially from the measured flat luminosity density evolution.
Combining our estimates with other measurements in the literature, we found $53\pm18$\% of the present stellar mass of early-type galaxies has been assembled at $z=0.7$.
\end{abstract}


\section{Introduction}
The luminosity functions (LFs) of galaxies provide fundamental clues to the evolution of galaxies. 
Until recently, measurements of the galaxy LFs were largely confined to near-IR to UV wavelengths 
(e.g., Blanton et al. 2003; Brown et al.\ 2007; Faber et al.\ 2007; Cirasuolo et al.\ 2007 for 
recent results) mainly due to the observational difficulties of covering large areas in the mid-IR.
The early mid/far-IR studies of galaxies utilized the \iras, \iso, and \cobe\ satellites and ground-based sub-millimeter observations.  These early studies led to the discovery of strong far-IR background radiation (Puget et al. 1996; Hauser et al. 1998; Fixsen et al. 1998), understanding the properties of luminous and ultra-luminous IR galaxies (e.g., Soifer et al. 1987; Sanders \& Mirabel 1996; Barger et al. 1999), and earlier measurements of LFs in the mid/far-IR bands (e.g., Soifer et al. 1987; Rowan-Robinson et al. 1987; Saunders et al. 1990;  Xu et al. 1998).
The \emph{Spitzer Space Telescope} (Werner et al. 2004), with its unprecedented capabilities, allows us to significantly expand on these results.
In particular,
the \spitzer/IRAC Channels 1--4 cover wavelengths of 3.6, 4.5, 5.8, and 8.0 $\mu$m (Fazio et al. 2004), 
respectively, providing unique windows for studies of galaxy properties.
At low redshifts ($z\simlt1$), the 3.6 $\mu$m and 4.5 $\mu$m channels lie on the Rayleigh-Jeans tail of the 
blackbody spectrum of stars, directly tracing the stellar mass
with little sensitivity to the ISM either through absorption or emission.
The 8.0 $\mu$m channel contains polycyclic aromatic hydrocarbon (PAH) features
whose luminosity is well correlated with star formation rates.
The situation for the 5.8 $\mu$m is more complicated since the galaxy flux is composed of mixture 
of starlight and PAH features.
Overall, the \spitzer/IRAC channels provide a comprehensive view of galaxy physics in the mid-infrared.

Recently, several studies have measured luminosity functions in the near and mid-IR based on IRAC photometry
in the Chandra Deep Field South (CDFS), Hubble Deep Field North (HDFN), Great Observatories Origins Deep Survey (GOODS), COMBO-17, and \spitzer\ Wide-Area Infrared Extra-galactic (SWIRE) surveys, where the majority of the galaxy redshifts are photometric redshifts,
complemented by spectroscopic redshifts from these surveys and the VIMOS VLT Deep Survey (VVDS, Le F\`evre et al. 2005).
In particular, there are studies using 2600 24\um\ sources with 1941 spectroscopic redshifts (Le Floc'h et al. 2005), 8000 24\um\ sources with photometric redshifts (P{\'e}rez-Gonz{\'a}lez et al. 2005), 1478 3.6\um\ sources with 47\% spectroscopic redshifts (Franceschini et al. 2006), 1349 24\um\ sources with photometric redshifts (Caputi et al. 2007), 17300--88600 3.6--24\um\ sources with photometric redshifts (SWIRE, Babbedge et al. 2006), and 21200 3.6\um\ sources with 1500 spectroscopic redshifts (Arnouts et al. 2007).  
Most of these studies use a combination of several photometric and redshift surveys.
Semi-analytical models of galaxy evolution have also been developed to 
provide theoretical basis for comparisons to the observed LFs (e.g., Lacey et al. 2008).

One of the central questions involved in these studies is the assembly history of galaxies,
which should depend on both galaxy type and luminosity.  
For example, Franceschini et al. (2006) found that the most massive galaxies are in place at $z\sim1$, 
while, including fainter galaxies, Arnouts et al. (2007) found about 50\% of quiescent and 80\% of active galaxies are in place at $z\sim1$.
Accurate LF measurements as a function of redshift are essential to such studies.
Unfortunately, dividing samples into redshift bins reduces the sample size and 
increases both statistical and systematic uncertainties.
The situation is further complicated by the presence of cosmic variance both within and between surveys and
the heavy dependence on photometric redshifts.  While photometric redshifts are frequently necessary for achieving 
a large sample size or a longer luminosity baseline, they can also lead to systematic uncertainties in the
shape and evolution of the luminosity functions.
Moreover, as we find in this study, the redshift dependence of the definition of luminosity can be a 
problem for estimates of evolution rates.

In this paper, we present mid-infrared galaxy luminosity functions for $z<0.6$ in the \spitzer/IRAC bands 
by combining the IRAC Shallow Survey (Eisenhardt et al. 2004) of the NOAO Deep Wide-Field Survey 
(NDWFS, Jannuzi \& Dey 1999) with redshifts from the AGN and Galaxy Evolution Survey 
(AGES, Kochanek et al. 2008, in preparation).
For the 3.6--8.0\um\ IRAC bands we have well-defined samples with roughly
 3800--5800 spectroscopic redshifts and a statistical power corresponding to samples of 4600--8000 objects through the use of random sparse sampling.
We describe the sample selection, photometry, and redshifts in \S2, the LF measurement methods in \S3, and the LFs in \S4.
We discuss our results in \S5. 
We assume that $H_0 = 100h~\rm{km~s^{-1}~Mpc^{-1}}$, $\Omega_{\rm m} = 0.3$, 
and $\Omega_{\Lambda}= 0.7$, and use the Vega magnitude system throughout the paper.

\section{Sample Selection\label{sec:sample}}
We measured the mid-infrared galaxy LFs in the NDWFS Bo\"otes field by combining
IRAC photometry from the IRAC Shallow Survey and AGES redshifts.
This field is also covered by multi-wavelength data in the UV (GALEX, Martin et al. 2005), 
optical (NDWFS), z-band (zBo\"otes, Cool 2007), 
near-IR (NDWFS and FLAMEX, Elston et al. 2006), and far-IR (24 $\mu$m, Soifer et al. 2004),
allowing us to use SED models to type the galaxies and make K-corrections (\S\ref{sec:sed}).
For simplicity, we briefly discuss the IRAC photometry and leave the discussion of photometry in the other 
bands to the references for each survey.

The IRAC Shallow Survey imaged the NDWFS field to $5\sigma$ limits of 18.4 (17.3), 17.7 (15.4), 15.5 (76)
and 14.8 (76) mag ($\mu Jy$) at [3.6]--[8.0] respectively for a 6\arcsec\ diameter aperture (Eisenhardt et al. 2004).
We used the IRAC Shallow Survey \verb+SExtractor+\footnote{http://terapix.iap.fr/soft/sextractor/.}
 \verb+MAG_AUTO+ (similar to Kron magnitudes)
and 6\arcsec\ aperture magnitudes for our analysis.
We note that the 6\arcsec\ aperture magnitudes include PSF corrections for flux losses outside the aperture.
Galaxies were selected as extended sources in the NDWFS optical data.  
We were concerned about the reliability of the total (from \verb+MAG_AUTO+) mid-IR magnitudes 
since \verb+MAG_AUTO+ may underestimate the total flux for extended 
galaxies by using a photometric aperture that is too small (e.g., Graham \& Driver 2005; Brown et al. 2007).
The problems can occur when the field is crowded with many sources or the 
exposure is relatively shallow, and will result in galaxy size-dependent
corrections that may mimic redshift evolution (see Appendix A).  
Since the optical bands have deeper images of the field with better PSFs, and 
the fixed aperture magnitudes do not suffer from this issue, we calculated the 
total$-$6\arcsec\ aperture magnitudes in the optical and in the IRAC bands to
test whether the IRAC photometry showed signs of such problem.
We found a difference of 0.15~mag between 
the total$-$6\arcsec\ aperture magnitudes in the optical and IRAC bands over the redshift range from 0 to 0.5.
In particular, at low redshift ($z=0$) the IRAC aperture correction is smaller than the optical bands.
Since we view the optical correction as more reliable, due to the higher resolution and 
greater relative depth of the data at these redshifts, we define the total magnitudes as the IRAC 6\arcsec\ aperture magnitude
corrected by the $R$-band total to 6\arcsec\ aperture correction,
\begin{equation}
m_{IRAC, total} \equiv m_{IRAC, 6^{\prime\prime}} + m_{R, total} - m_{R, 6^{\prime\prime}}.
\end{equation}
There may be further uncertainties in the \verb+MAG_AUTO+ $R$-band magnitudes, 
but we expect they affect our estimates of 
 the galaxy evolution rate by less than our statistical uncertainties (\S\ref{sec:levo} and Appendix A).

\begin{figure}
\epsscale{0.60}
\plotone{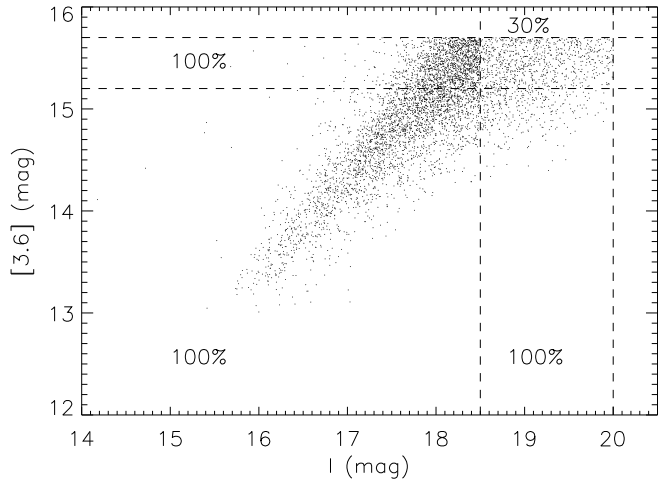}
\caption{[3.6] spectroscopic selection.  The points show the distribution of galaxies with measured redshifts.
The [3.6] sample is defined by all galaxies with $[3.6] \le 15.7$~mag and $I \le 20$~mag.
However, we only sample 30\% of galaxies in the range of $15.2 \le [3.6] \le 15.7$ and $18.5 \le I \le 20$.
The [4.5], [5.8], and [8.0] samples have similar definitions but with different magnitude limits.
\label{fig:iraci}}
\end{figure}

\begin{figure}
\epsscale{0.60}
\plotone{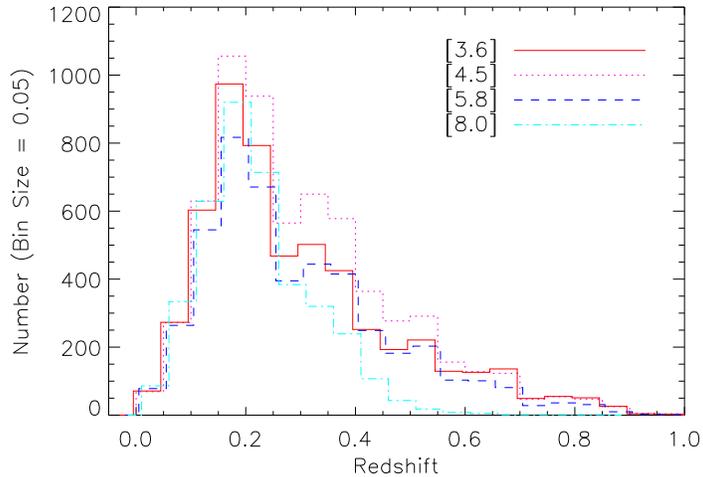}
\caption{The redshift distributions of our sample galaxies in the IRAC [3.6] to [8.0] bands.  
The histograms are
binned with $\Delta z = 0.05$, and are slightly offset ($< 0.01$) in redshift for clarity.
\label{fig:z}}
\end{figure}

The AGES redshift survey covers most of the NDWFS field.  
Spectra were obtained for well-defined samples of galaxies in all the NDWFS photometric bands
as well as for samples of AGNs. 
Our present sample combines limits from the AGES $I$-band and IRAC redshift sampling strategies (Table~\ref{tab:sample}).  
In the $I$-band, AGES targeted all galaxies with $I \le 18.5$~mag, complementing the 
$I<16$~mag part of the sample with redshifts from the SDSS (Adelman-McCarthy et al. 2007) survey,  
 and then randomly selected 20\% of galaxies with $18.5 \le I \le 20$~mag for redshifts.
 Based on the IRAC-optical magnitude distributions (see Figure~\ref{fig:iraci}), AGES chose $C1$--$C4$ ([3.6]--[8.0])
magnitude limits of $m_{lim}([X]) = $ 15.7, 15.7, 15.2, and 13.8~mag respectively.
AGES then attempted to obtain redshifts for all galaxies brighter than $m_{comp}([X]) =$ 15.2, 15.2, 14.7, and 13.2~mag for
[3.6]--[8.0] respectively, and a randomly selected 30\% of the galaxies in the magnitude ranges of 
$m_{comp}([X]) < [X] \leq m_{lim}([X])$.
Figure~\ref{fig:iraci} illustrates the combined effect of the optical and mid-IR sampling.
We have a sample weight of $f_s=1$ for $I \le 18.5$~mag or $[X] \le m_{comp}([X])$ and $f_s=0.3$ for $I\ge18.5$~mag and 
$m_{comp}([X]) \le [X] \le m_{lim}([X])$.
The redshift completenesses are
92\%, 93\%, 93\% and 96\% for the 3.6 to 8.0\um\ bands, with no significant variations in the completeness with magnitude.
Figure~\ref{fig:z} shows the measured redshift distributions.  
The distributions peak at $z\simeq0.2$ and extend to $z\sim1$, although we limit our analysis to $z\le0.6$.
We excluded 155 AGNs (see \S\ref{sec:sed}) from the analysis.
The final samples consist of 4905, 5847, 4367, and 3802 galaxies corresponding to a statistical
sample of 6111, 7826, 5499, and 4782 objects due to the random sparse sampling.

The spectra for the AGES survey were obtained with the 300 fiber Hectospec Spectrograph (Fabricant et al. 2005) on the 6.5m MMT.
Hectospec covers the wavelength range from 3200 to 9200\AA,
with a resolution of $R \simeq 1000$.  With multiple runs and passes
over the field, objects with initially poor spectra were systematically
re-observed in order to produce the final, high completeness of the
redshift samples.
They were reduced using both the Hectospec pipeline at the CFA and a modified SDSS pipeline 
(HSRED\footnote{http://mizar.as.arizona.edu/rcool/hsred/.}). 
The redshifts were verified to be correct by visual inspection.
Because there were multiple pointings for each region of the survey, fiber collision limits
in the individual pointings have little consequence for our sample completeness.
The survey area for the AGES main sample, including effects such as excluding regions close to bright stars, is 7.44 square degrees.

\begin{deluxetable}{cccccccc}
\tabletypesize{\scriptsize}
\tablecolumns{8}
\tablewidth{0pt}
\tablecaption{Sample Selection and Redshift Targeting Limits for IRAC Bands\label{tab:sample}}
\tablehead{
\colhead{Band} &
\colhead{$m_{comp}([X])$} &
\colhead{$m_{lim}([X])$} &
\colhead{$I_{comp}$} &
\colhead{$I_{lim}$} &
\colhead{Redshift} &
\colhead{Survey Area} & 
\colhead{Number of} \\
\colhead{} &
\colhead{(mag)} &
\colhead{(mag)} &
\colhead{(mag)} &
\colhead{(mag)} &
\colhead{Completeness} &
\colhead{(square degree)} &
\colhead{Galaxies}
}
\startdata
C1 (3.6\um) & 15.2 & 15.7 & 18.5 & 20 & 0.92 & 7.44 & 4905 \\
C2 (4.5\um) & 15.2 & 15.7 & 18.5 & 20 & 0.93 & 7.44 & 5847 \\
C3 (5.8\um) & 14.7 & 15.2 & 18.5 & 20 & 0.93 & 7.44 & 4367 \\
C4 (8.0\um) & 13.2 & 13.8 & 18.5 & 20 & 0.96 & 7.44 & 3802 \\
\enddata
\tablecomments{$m_{lim}([X])$ and $I_{lim}$ are the flux limits of the sample selection.
$m_{comp}([X])$ and $I_{comp}$ are the limits where 100\% galaxies brighter than these 
magnitudes are targeted for redshifts.  30\% of galaxies with 
$m_{comp}([X]) \le [X] \le m_{lim}([X])$ and
20\% of galaxies with $18.5 \le I \le 20$~mag are targeted for redshifts.}
\end{deluxetable}

\begin{figure}
\epsscale{0.8}
\plotone{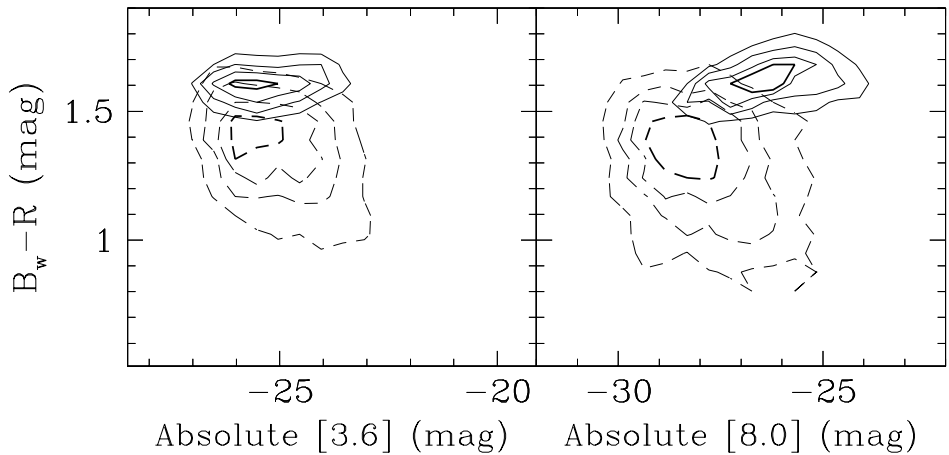}
\caption{The rest frame color ($B_W-R$) versus magnitude plot for the [3.6] and [8.0] bands.  
The contours enclose 20\% (bold), 40\%, 60\% and 80\% of
the objects with solid lines for early-type and dashed lines for late-type. 
We defined an early-type galaxy to be one in which 80\% or more 
of the 0.2--10\um\ luminosity is assigned to the early-type template of Assef et al.\ (2008).
Otherwise, we defined it a late-type galaxy.\label{fig:el}}
\end{figure}

\begin{figure}
\epsscale{1}
\plotone{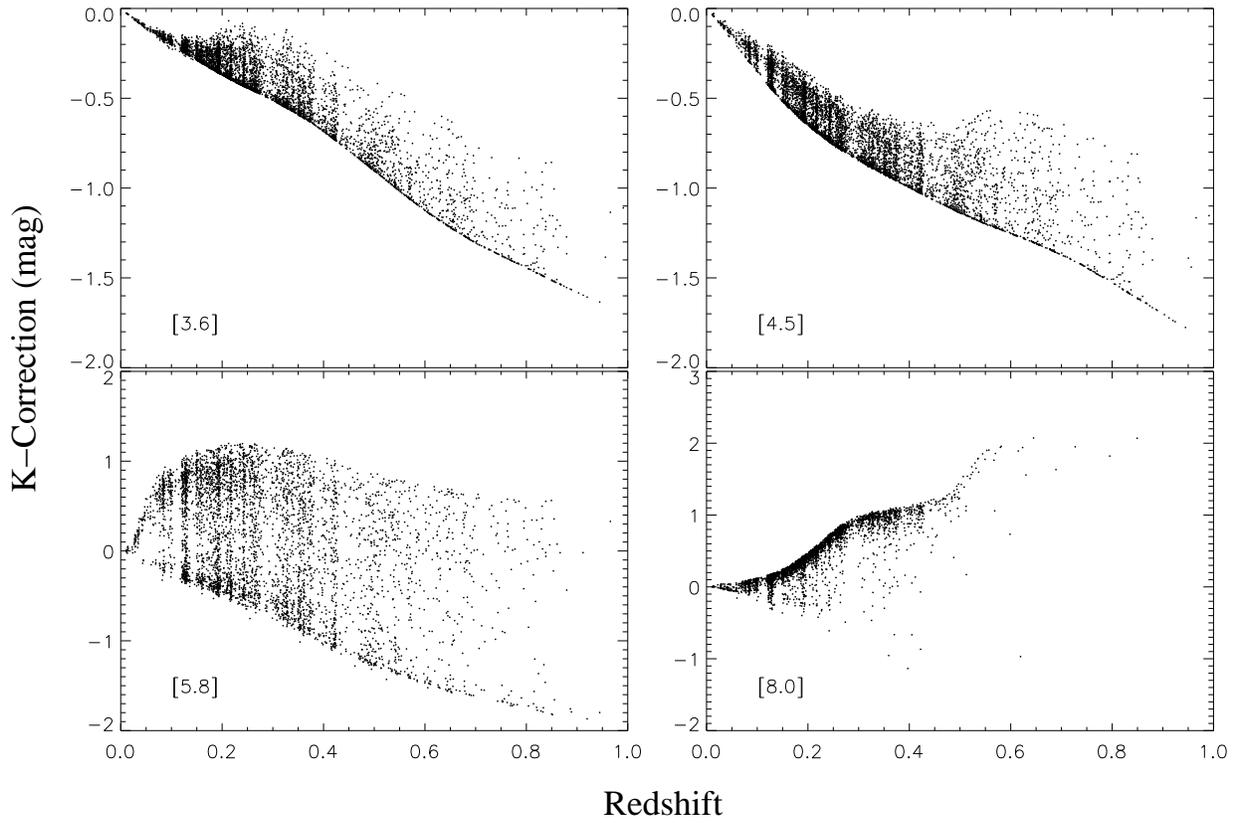}
\caption{The K-corrections for galaxies in the IRAC [3.6]--[8.0] bands, respectively.
The scatter of the K-corrections in the [8.0] band is smaller than that for the [5.8] band because 
relatively few early-type galaxies are detected in the [8.0] band.
\label{fig:kcor}}
\end{figure}

\begin{figure}
	\epsscale{0.9}
	\plotone{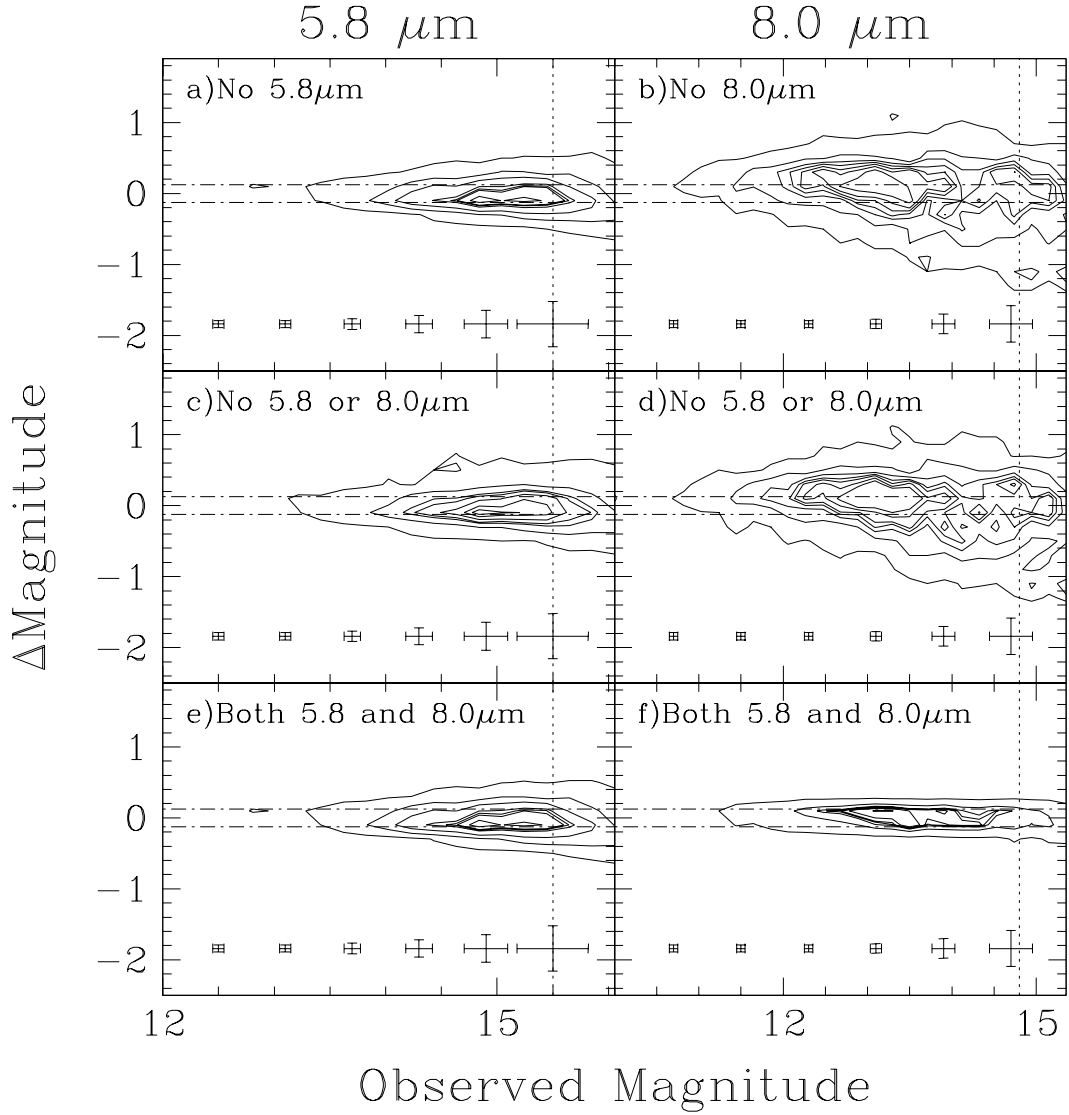}
	\caption{Contours of the magnitude
	difference between the model and measured [5.8] and [8.0] 
	magnitudes for the three cases.  Case 1, ``both [5.8] and [8.0]'' (bottom) uses all the photometric data. 
	Case 2, ``no [5.8] or [8.0]'' (middle) uses neither the 5.8 nor the 
	8.0 micron data.  Case 3 ``no [5.8]'' for the 5.8\um\
	comparison, and ``no [8.0]'' for the 8.0\um\ comparison (top) uses all bands but without 
	the one we are considering.
	The left column is for the [5.8] 
	channel and the right for the [8.0] channel.
	The density contours encompass
	90, 70, 50, 40, 30, 20, and 10\% of the galaxies.  The horizontal
	lines show the width of one of our magnitude bins, and the vertical
	line shows the magnitude limit used for the band.  A sequence of
	error bars at the bottom indicate the average uncertainties as
	a function of magnitude.  
	\label{fig:kcora}}
\end{figure}

\begin{figure}
	\epsscale{0.9}
	\plotone{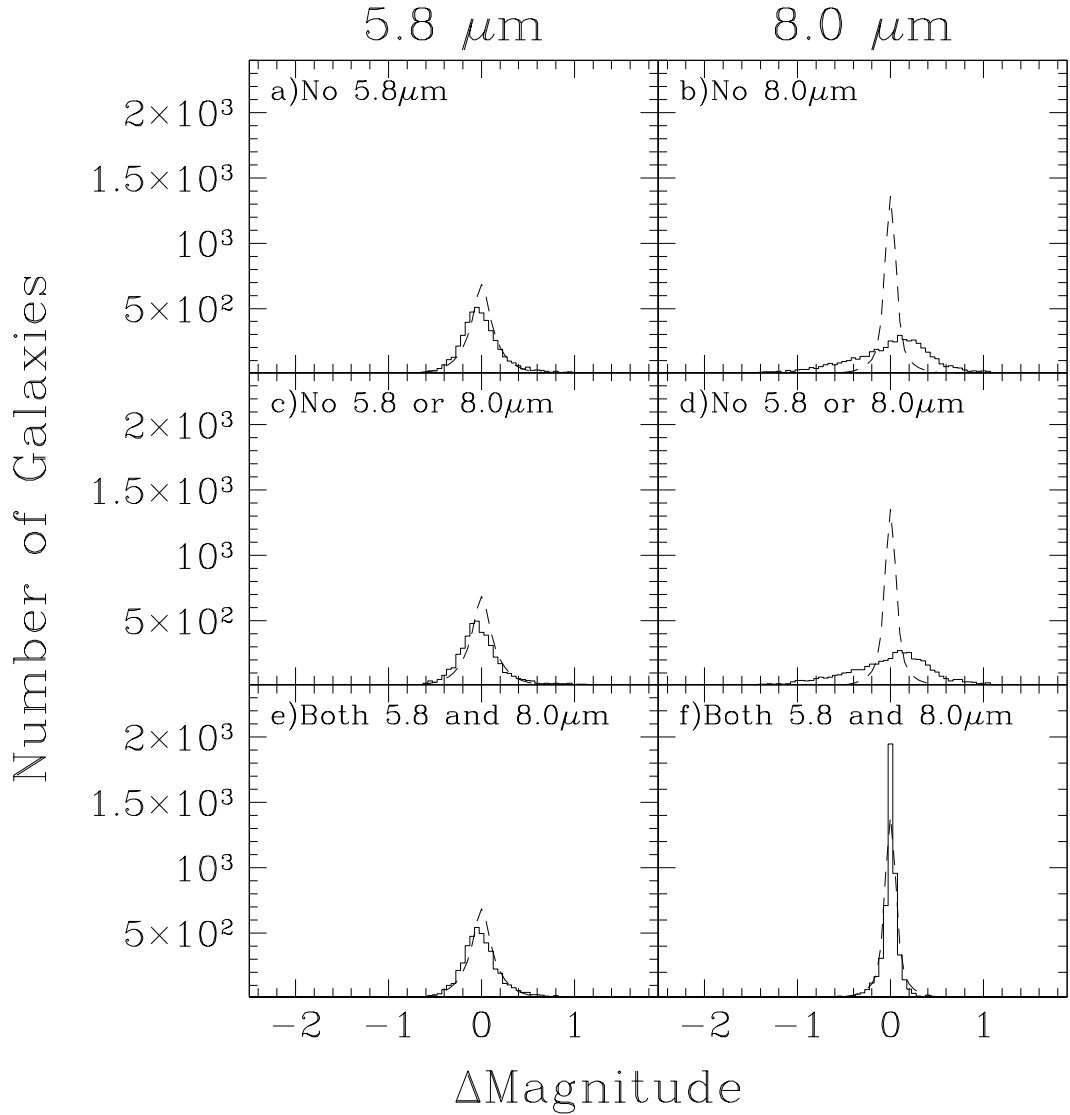}
	\caption{Histograms of the magnitude
	differences calculated in Figure~\ref{fig:kcora} for galaxies above the magnitude limits (solid) as
	compared to the distribution we would expect given the uncertainties
	in the magnitudes (dashed).
	\label{fig:kcorb}}
\end{figure}

\subsection{SED Modeling\label{sec:sed}}
We fit the $B_W$, $R$, $I$, $z$, $J$, $K_s$, $K$, [3.6], [4.5], [5.8] and [8.0] 6\arcsec\ diameter aperture magnitudes for each 
galaxy with combinations of spectral templates for early, late, and irregular galaxies developed 
by Assef et al.\ (2008) based on the same data.
The choice of the 6\arcsec\ aperture represents a compromise between smaller apertures that are more sensitive to aperture corrections and
larger apertures that are more sensitive to contamination by
other sources.
Because of the magnitude limits required for the spectroscopy, 
the galaxies all have photometry in at least 4 of these bands, 
and on average have measurements in 9 bands.  
From the Assef et al.\ (2008) models, 
we fit all available data (0.2--10\um) with
the early, late, and irregular templates for each galaxy and defined an early-type galaxy to be one in which 80\% or more 
of the total 0.2--10\um\ luminosity is assigned to the early-type template.
The 80\% value is at the minimum of the bimodal distribution
of early-type fractions (see Assef et al.\ 2008).
The late-type galaxies are defined to be those do not satisfy the early-type criterion.  
This separates the ``red sequence'' from the ``blue cloud'', as we show in Figure~\ref{fig:el}.
After obtaining the spectral model for each galaxy, we used the templates (Assef et al.\ 2008)
to compute the K-corrections that are needed for determining rest frame luminosities.
These K-corrections are consistent with the analytical approximations in Huang et al. (2007), as we show in Figure~\ref{fig:kcor}.  
The large scatter of the K-corrections in the [5.8] band is due to the spectral differences between the early and late-type galaxies, 
where the PAH features contribute significantly to the late-type galaxies but not the early-type galaxies.  
Although the [8.0] band is dominated by the PAH emission in late-type galaxies, 
the scatter in the K-corrections is not as large as that in the [5.8] band 
because relatively few early-type galaxies are detected in the [8.0] band.
The spectral fitting process also enables us to identify 155 galaxies that have significant AGN flux 
contributions (reduced $\chi^2 > 50$ in the template fits), which are then excluded for determining the LFs. 
These galaxies generally show the flat mid-IR power-laws characteristic of AGNs (Stern et al. 2005).

We test the robustness of the SED model in the [5.8] and [8.0] bands.
While we cannot directly test our estimate of the rest frame magnitudes,
we can examine how well the models reproduce the observed frame magnitudes
as we use less information.  In particular, we can fit the photometric
data by dropping one or both of the [5.8] or [8.0] data points and then compare
the predicted and measured magnitudes.  This is a somewhat unfair 
comparison, of course, because for the real sample we always possess
the [5.8] and [8.0] magnitudes.
We carried out the experiment for three cases.  The first
case is simply how well the SED models fit the [5.8] and [8.0] data
using the fits to all the photometric data (``both 5.8 and 8.0'').
In the second case we fit the SED using neither the [5.8] nor the 
[8.0] data (``no 5.8 or 8.0'').  Finally, in the third case
we drop only the band we are considering (``no 5.8'' for the [5.8]
comparison, and ``no 8.0'' for the [8.0] comparison).
Figure~\ref{fig:kcora} shows contours of the magnitude
difference between the model and measured magnitudes.  
Figure~\ref{fig:kcorb} shows histograms of the magnitude
differences for galaxies above the magnitude limits as
compared to the distribution we would expect given the uncertainties
in the magnitudes.  

Based on these two plots, we can see that the K-corrections for the 
[5.8] channel are very robust.  Even if we use neither the [5.8] nor the
[8.0] magnitudes, the distribution of the magnitude differences is modest
compared to our luminosity function bin widths and are largely consistent with the measurement
uncertainties.  This holds at 8 microns when we use all the available
data, but there is significant scatter if we do not use the 5.8 and/or
8 micron data.  Nonetheless, while it is broader than our magnitude
bins, it would affect our results little.
After all, we always have the 5.8 and 8.0 micron data, and so we 
are generally closer to the ``both 5.8 and 8.0''
micron case than to the others.  The redshifts of the 8.0 micron 
samples remain low enough that the 8.0 micron band generally samples
portions of the PAH emission and so should remain well-behaved.
The worst case scenario is that at $z\sim 0.4$ we are approaching the 
``no 8.0 micron'' case for the 8.0 micron K-corrections.

\section{Luminosity Function Determining Methods\label{sec:method}}
We used both the parametric maximum-likelihood method (STY, Sandage et al.\ 1979) 
and the non-parametric stepwise maximum-likelihood method (SWML, Efstathiou et al.\ 1988) to fit the luminosity functions.  
In the STY method, we parameterized the LF as a Schechter function
\begin{equation}
\Phi(M)dM = (0.4\ln10)\phi_*(10^{0.4(M_*-M)})^{1+\alpha}\exp(-10^{0.4(M_*-M)})dM
\end{equation}
and we allowed $M_*$ to evolve as 
\begin{equation}
M_*(z) = M_*(0) - Qz
\end{equation}
when fitting the LF, following the parameterization of Lin et al. (1999).  
We did not allow the normalization $\phi_*$ or the slope $\alpha$ of the LF to evolve since our sample size is not large 
enough to constrain the evolution of these parameters.
Estimates of $\phi_*$ will also suffer from cosmic variance at low redshifts where we 
have little volume.
In the SWML method, the LF is defined in bins $M_k - \Delta M/2 \le M \le M_k+\Delta M/2$ with value $\Phi_k$.
In both methods, the parameters of the LFs in the STY method ($M_*$, $\alpha$, and $Q$) and the SWML ($\Phi_k$) method are calculated by maximizing the likelihood functions.
Since the absolute normalization factor $\phi_*$ is not modeled in the likelihood function, the shape of the LF determined from the STY and SWML methods is not
sensitive to the effects of large scale structure.
We calculated the normalization $\phi_*$ using the minimum variance method (David \& Huchra 1982).  Since the STY and minimum variance methods are widely used in determinations of LFs, we leave the details of the calculations to other references (e.g., Lin et al. 1996, 1999).
We checked the STY and SWML calculations using both synthetic catalogs and by calculating
 the LFs using the $1/V_{max}$ method (Schmidt 1968; Avni \& Bahcall 1980).  The LFs from the $1/V_{max}$ method are in very good agreement with those 
determined from the SWML method, except for the very low luminosity bins where the $1/V_{max}$ results show a deficit of galaxies 
compared to the SWML method, probably due to the effects of large scale structure.  We only present the STY/SWML results.

Each galaxy was assigned a weight based on the sampling strategy and the redshift completeness.  
The mean redshift completeness, $f_z = 0.92$, 
0.93, 0.93 and 0.96 for the [3.6]--[8.0] bands, depends little on magnitude, and the sampling fraction $f_s = 1$ or 0.3 depends on the
target magnitudes and the band  (see \S2.2 Table~1 and Figure~\ref{fig:iraci}).  Thus, each galaxy has an overall statistical weight of $(f_zf_s)^{-1}$.
We included both the IRAC and $I$ band selection limits in our LF measurements.
We also carried out the calculations using only the IRAC selection limits, and the resulting LFs are consistent with the full analysis.
This is expected, since at the IRAC magnitude limits we are losing few galaxies due to the optical flux limit (see Figure~\ref{fig:iraci}).

We measured the LFs using the STY and SWML methods in three different ways for the [3.6]--[8.0] bands.
First, we determine the LFs by applying the STY method to the entire redshift range ($z<0.6$).
This is essentially a pure luminosity evolution model since we did not allow the normalization $\phi_*$ 
or the slope $\alpha$ of the LF to evolve.
Second, we applied the STY method to the three redshift bins $z<0.2$, $0.2<z<0.35$, and $0.35<z<0.6$. 
We chose the redshift bins as a balance between the number of galaxies (statistical uncertainties) and the bin width (averaging over cosmic time).
We fixed the slope $\alpha$ and evolution parameter $Q$ to the values obtained from the first method and fit only
the normalization $\phi_*$ and $M_*$ in each redshift bin.
This allows us to check whether the LFs evolve beyond the pure luminosity evolution model.
Third, we measured the binned LFs with the SWML method in the three redshift bins and then fit them
jointly with Schechter functions, where we fixed the faint end slope $\alpha$ to be the same in all bins but allowed
the $M_*$ and the normalization $\phi_*$ to differ.

\section{Luminosity Functions}

\begin{figure}
\epsscale{1}
\plotone{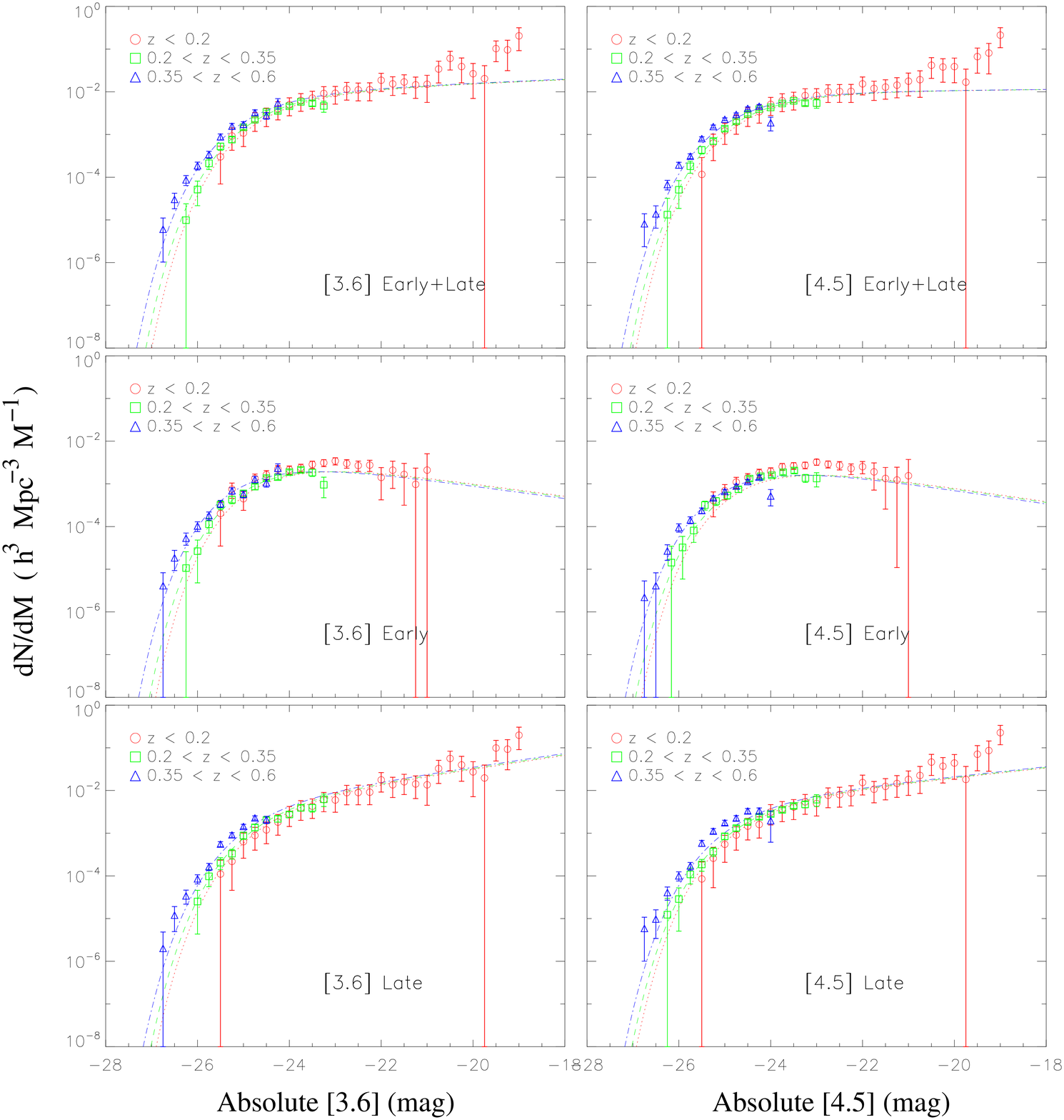}
\caption{The luminosity functions in the [3.6] and [4.5] bands for total, early, and late-type galaxies.  The LFs are plotted in three redshift bins with different colors.  The lines are the LFs determined from the STY method fitting to the whole sample but plotted in different redshift bins.  The discrete points are the LFs from the SWML method.  The LFs from the STY and SWML method are 
generally consistent with each other indicating our pure luminosity evolution
parameterization in the STY method is reasonable.  
\label{fig:cone}}
\end{figure}

\begin{figure}
\epsscale{1}
\plotone{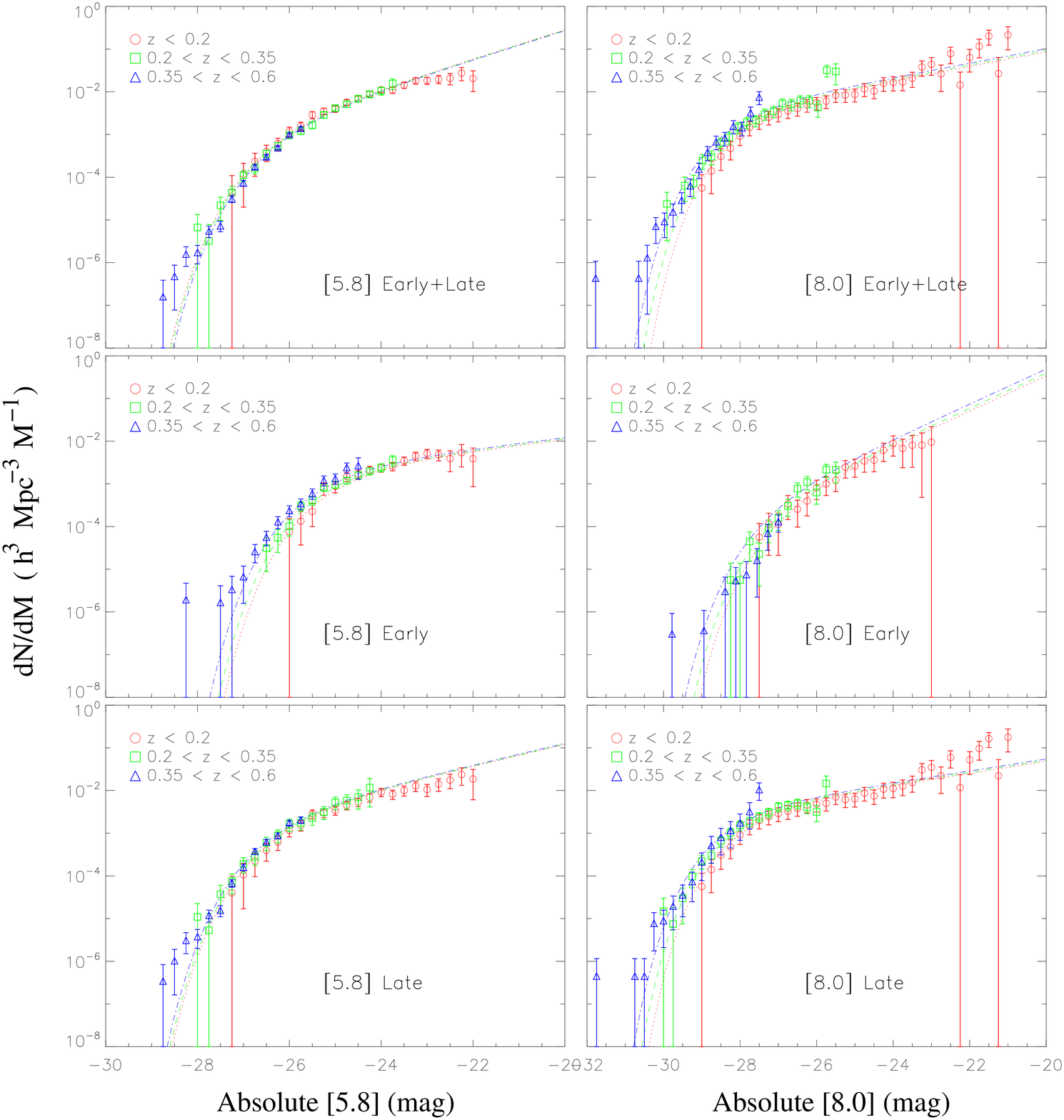}
\caption{The luminosity functions in the [5.8] and [8.0] bands with same sets of symbols/lines as in Figure~\ref{fig:cone}.  
The LFs from the STY and SWML method are generally consistent with each other
indicating pure luminosity evolution.
\label{fig:cthree}}
\end{figure}

Figures~\ref{fig:cone} and \ref{fig:cthree} show the resulting luminosity functions for the
four bands.  For each band we show the early-type, late-type and total LFs for redshift
bins of $0<z<0.2$, $0.2<z<0.35$ and $0.35<z<0.6$.  We show only the non-parametric
SWML LFs and the result for the global parametric STY fit evaluated at
the median of the redshift bin.  Table~\ref{tab:styfit} presents the parameters for the global STY fits 
with an evolving $M_*$, Table~\ref{tab:swmlfit} presents the Schechter function fits to the SWML fits 
for the three redshift bins, and Table~\ref{tab:swml} presents the tabulated SWML LFs.  
The STY and SWML
estimates of the luminosity functions are broadly consistent with each other, although 
we do find systematic mismatches in some cases (such as the faint end slope of the 
total LF in the [5.8] band with 1--2$\sigma$ offsets).   
Given the flux limits of our survey, we can only determine the full LF parameters in
the global fits and the lowest redshift bin.  For the higher redshift bins we only 
sample the higher luminosity galaxies and cannot reliably determine the faint-end
slope $\alpha$.
The SWML LFs are well fit by the Schechter functions with $\chi^2/dof < 1$ in all cases 
(Table~\ref{tab:swmlfit}), which also suggest that the SWML method slightly over-estimates
the error-bars on the binned LFs.

We can compare our results to several recent mid-IR luminosity function measurements
based on Spitzer data.  
Arnouts et al. (2007) and Babbedge et al. (2006) derived mid-IR luminosity 
functions based on the Spitzer Wide-Area Infrared Survey (SWIRE).  Babbedge et al.
(2006) used the 6.5~deg$^2$ ELAIS-N1 field of the SWIRE survey based on photometric redshifts
and using the $V/V_{max}$ method to derive the luminosity functions based on 
approximately 34000, 34000, 14000 and 17000 galaxies to magnitude limits of
$17.4$ ($30\mu$Jy) , $16.9$ ($30\mu$Jy) , $15.9$ ($50\mu$Jy) and $15.3$ ($50\mu$Jy) 
for the [3.6]--[8.0] bands respectively.  Arnouts et al. (2007) used the SWIRE data
for the $0.85$~deg$^2$ VVDS--0226--04 field using a combination of spectroscopic
and photometric redshifts with limiting [3.6] magnitudes of $17.7$ and $18.2$ 
respectively.  They derived a rest-frame K-band luminosity function using the
$V/V_{max}$ and STY methods.  The number of galaxies used in each analysis is
unclear, but 1500 redshifts were available for the field.  Huang et al. (2007)
analyzed a subsample at [8.0] of the present data to 13.5~mag at
$z<0.3$ to estimate the local [8.0] luminosity function.     
We will convert from K-band back to the [3.6] band using
the median rest-frame color of K--[3.6]=0.41~mag found for the $z<0.35$  
AGES galaxies.

\begin{figure}
\epsscale{1}
\plotone{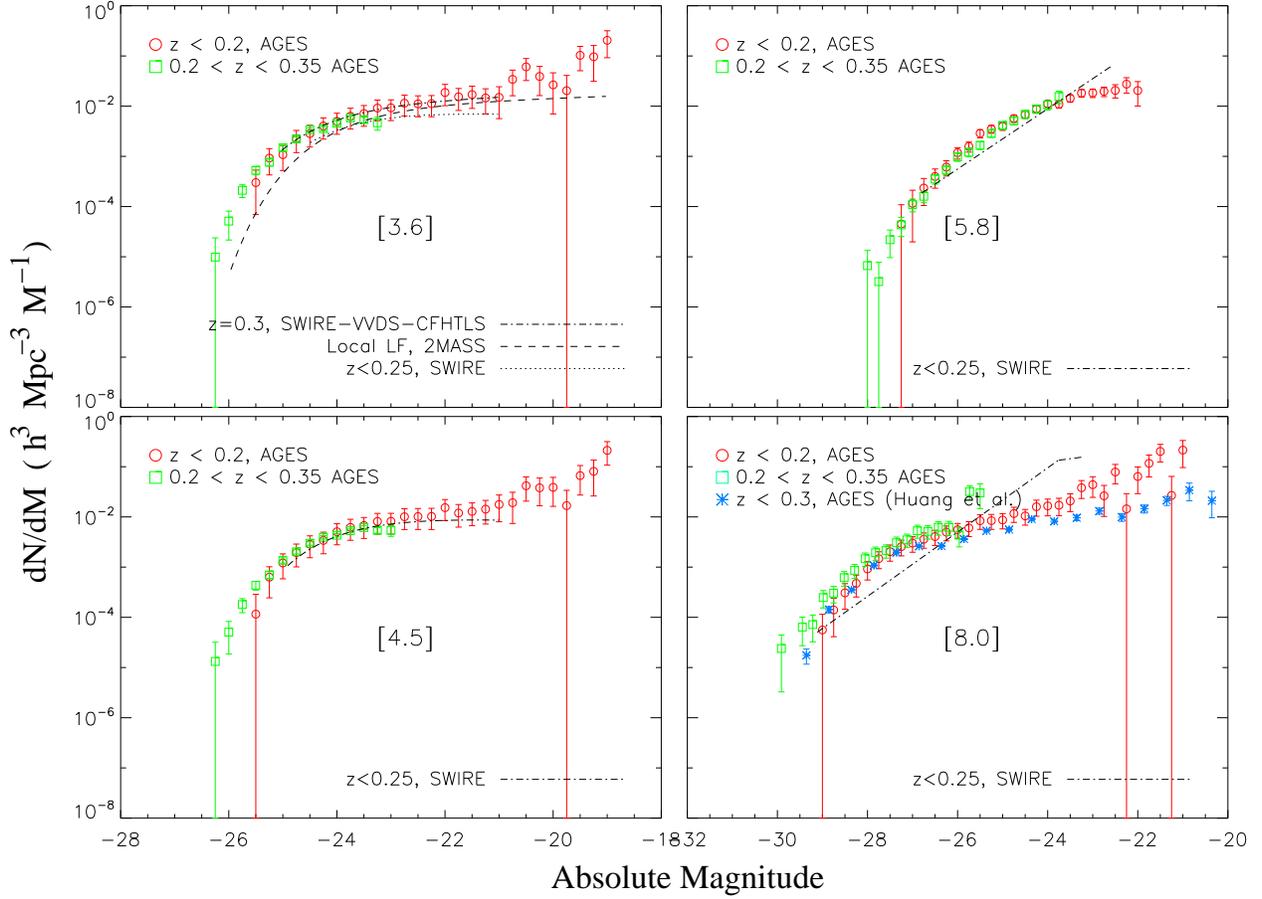}
\caption{Comparisons between the [3.6] through [8.0] LFs from AGES at $z<0.25$ and $0.2<z<0.35$ and those from SWIRE/VVDS/CFHTLS (Arnouts et al. 2007), 2MASS (Kochanek et al. 2001), SWIRE (Babbedge et al. 2006), and Huang et al. (2007) results after correcting for the color, cosmology, and magnitude system differences.
There is generally good agreement between our LFs and other measurements,
except in the [5.8] and [8.0] bands where Babbedge et al. (2006) found that the LFs were
better fit by power-laws.
\label{fig:com}}
\end{figure}

Fig.~\ref{fig:com} presents the comparisons both for these Spitzer samples and the 2MASS
K-band luminosity function of Kochanek et al. (2001) for the total LF.  
We only show the 
parametric fits to the comparison luminosity functions over the magnitude
range for which the other survey had data, and converted their results to
Vega magnitudes and our cosmological model as necessary.  
These are summarized in Table~\ref{tab:comp}.
At [3.6] and [4.5], our 
results agree well with the other luminosity functions.  In particular,
at [3.6], all three parameters for the $z<0.2$ bin are consistent with 
the local 2MASS values, although a small magnitude shift may be present,
indicative of the luminosity evolution or luminosity definition problems we discuss in \S\ref{sec:levo} and Appendix A.  
Our faint end slopes in [3.6] and [4.5] for the total LF, 
$\alpha=-1.12\pm0.16$ and $-1.01\pm0.13$, 
are consistent with that in 2MASS, $-1.09\pm0.06$ (Kochanek et al. 2001).
Parameter comparisons with Arnouts et al. (2007) and
Babbedge et al. (2006) are somewhat moot due to their larger uncertainties.
At [5.8] and [8.0] we can only compare to Babbedge et al. (2006) and
our own earlier result in Huang et al. (2007).  Huang et al. (2007)
used a different sample definition and analysis method but was based 
on the same photometric and redshift surveys, and it is not surprising that our results
are in agreement, except for the faint end tail (see Fig.~\ref{fig:com}).  
We do not agree with the general structure of the
[5.8] and [8.0] LFs found by Babbedge et al. (2006), who found a better fit using 
power laws at the bright end rather than having the 
exponential truncation of the Schechter function.  We see some very
weak evidence for such an extension at [5.8] but no
evidence for a global, bright-end power-law. 

\subsection{Comparison between Early and Late-Type Galaxies}
In the [3.6], [4.5], and [5.8] bands we find that the early-type galaxies have shallower
faint end slopes than the later-type galaxies, in agreement with Arnouts et al. (2007), 
who also separated the early and late-type galaxies using SED models.
The situation reverses at [8.0] with the early-type showing a steeper faint end slope,
with 1.4$\sigma$ confidence, than the late-type's.
In particular in the [3.6] and [4.5] bands, our faint end slopes for late-type galaxies,
$\alpha_l=-1.40\pm0.18$ and $-1.29\pm0.17$, 
are consistent with the value, $\alpha_l=-1.3\pm0.2$, from Arnouts et al. (2007), while in early-type galaxies there is about $1\sigma$ difference between our estimate ($\alpha_e=-0.6\pm0.3$) and Arnouts et al.'s ($\alpha_e=-0.3\pm0.2$).
Our slopes are steeper than what Kochanek et al. (2001) found for their
morphologically typed samples, $\alpha_e=-0.92\pm0.10$ (1$\sigma$ difference) and $\alpha_l=-0.87\pm0.09$ (2.6$\sigma$ difference), but this could be due to the different 
type definitions.  Certainly, our present criterion of defining type 
based on a fixed luminosity fraction from the early-type template will tend to 
make the faint end slope less negative because low luminosity early-type
galaxies are on average optically bluer (e.g., Figure 6 of Brown et al.\ 2007) and will hence be shifted towards the
type boundary.  The early and late-type galaxies do show a systematic
difference in their K--[3.6] colors, with values of 0.35 and 0.54~mag for the
median rest frame color of the early and late-type galaxies respectively.
The shift for the late-type galaxies is presumably due to a larger PAH
contribution in [3.6] for late-type galaxies, and it helps to explain
why the $M_*$ of early and late-type galaxies at [3.6] are more
similar than they are at K-band.
In addition, the differences between the faint end slopes in this
paper and Kochanek et al. (2001) also affect the estimates of $M_*$. 
The absence of strong PAH emission in the
early-type galaxies leads to dramatic differences in the [8.0] 
band -- the early-type galaxies are significantly fainter than the
late-type galaxies, and they show a very steep faint-end slope.

\subsection{Luminosity Evolution \label{sec:levo}}

\begin{figure}
\epsscale{1}
\plotone{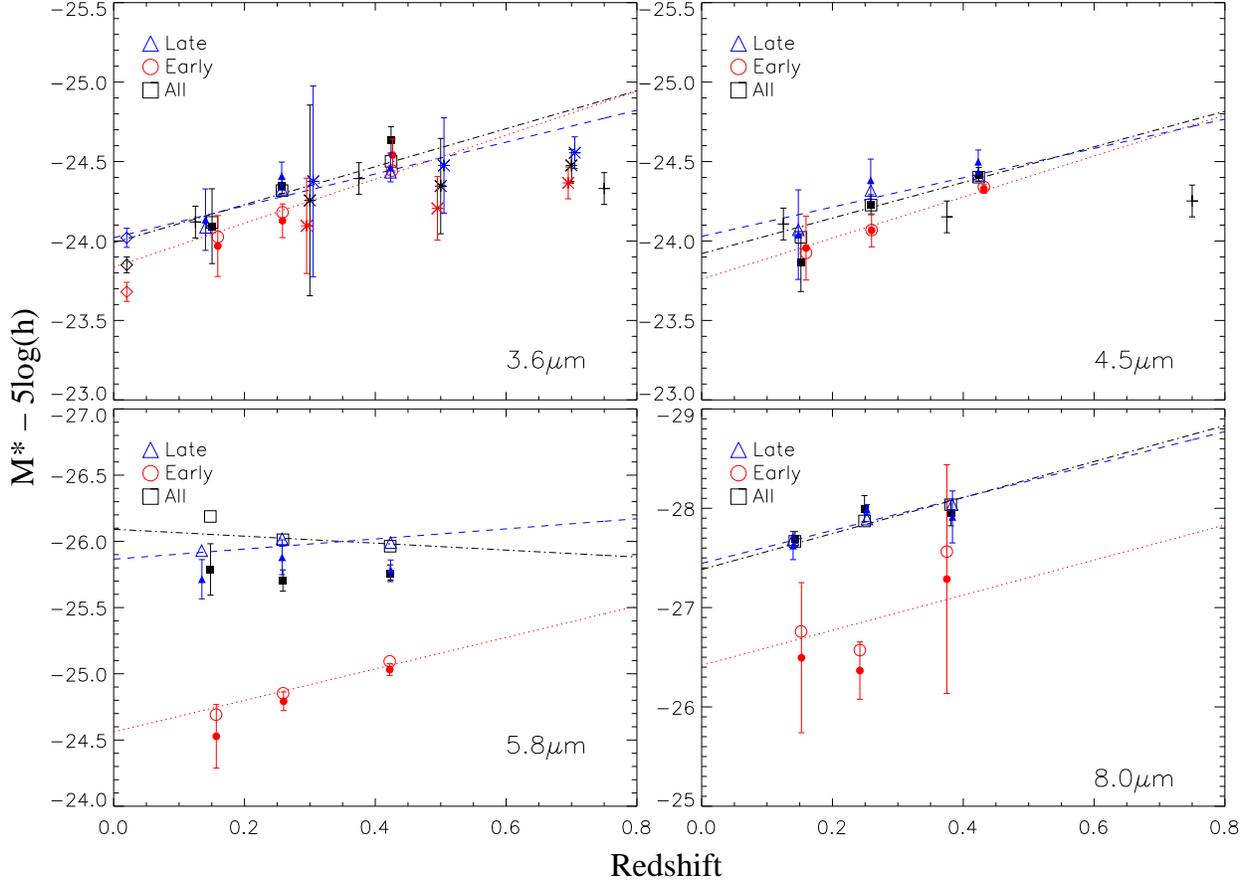}
\caption{The evolution of $M_*$ in the IRAC channels for the total (black, square, dot-dashed), 
early (red, circle, dotted) and late (blue, triangle, dashed) type galaxies.  
The lines are the results from the STY fit to the full sample.  
The triangles, circles and squares are our AGES results, where the filled symbols are the results 
from fitting Schechter functions to the binned SWML LFs, and the open symbols are the results from 
the STY fits to the individual redshift bins.  We do not plot the error-bars from the STY 
fits in the individual redshift bins for clarity, 
but they are similar to those determined from the SWML LFs.
The diamonds, stars, and ``+'' signs are results from Kochanek et al. (2001), Arnouts et al. (2007) and Babbedge et al. (2006) after correcting for color, cosmology, magnitude system, and faint end slope differences.
In the [3.6] and [4.5] bands, which trace the Rayleigh-Jeans tail of the stellar emission, 
the evolution of $M_*$ is consistent with a passive evolution model, while in the [8.0] band, 
which is sensitive to star formation, the evolution of $M_*$ 
is consistent with other estimates for the evolution of star formation rates.
\label{fig:mstar}}
\end{figure}

Figure~\ref{fig:mstar} shows the evolution of $M_*$ with redshift.  We have
three estimates of the evolution.  One from the values of $Q$ derived in the global
STY fits, one from the STY fits to the individual bins, and one from the
Schechter function fits to the SWML luminosity functions.  The differences
between the three estimates are generally smaller than the statistical
uncertainties, suggesting that our global STY fits are using an acceptable
parametrization of the evolution.  Where there are differences between the
results, they are generally due to shifts in the faint end slope $\alpha$ between
the various fits.  For example, the differences in the [5.8] bands are due
to the SWML fits giving a shallower slope than the STY fits ($\alpha = -1.60\pm0.07$
versus $-1.85\pm0.13$).
  
If we transform the Kochanek et al. (2001)
K-band points to [3.6], they lie on the redshift zero extrapolation of
our [3.6] results, and if we transform the Arnouts et al. (2007) results
back to [3.6] they also lie on our trend.  
The $M_*$ values from Babbedge et al. (2006) are also consistent with our trend
for the [3.6] and [4.5].
Since $M_*$ is correlated with the faint end slope, we corrected for the correlation
in this comparison by 
using the $M_*$--$\alpha$ confidence contours from our measurements
to shift the other results for $M_*$ to match our estimates of $\alpha$.
Because of the very different power-law form used by Babbedge et al. (2006)
at [5.8] and [8.0] we cannot compare to their results in the longer wavelength
bands.      

At [3.6] and [4.5] the evolution rates depend little on galaxy type,
both in our results and in the earlier studies.  The behavior is very
different at [5.8].  At [5.8] we see essentially no 
evolution for the late-type galaxies and a steady brightening of the
early-type galaxies. 
The enormous uncertainty in early-type galaxies at [8.0] does not allow
us to perform a meaningful comparison.

The [8.0] band should trace the star formation rate through the emission from the PAH features.
Our LF evolution rate $Q=1.8\pm1.1$ in [8.0], equivalent to 
$\beta=2.4\pm1.5$ for $L_{SFR} \propto (1+z)^{\beta}$, is consistent with other 
estimates for the evolution of star formation (e.g., $\beta=2.7\pm0.6$, Hopkins 2004; 
$\beta= 3.8\pm0.5$, Villar et al. 2008).
The [3.6] and [4.5] bands largely trace the Rayleigh-Jeans tail of the stellar emission,
where the LFs are expected to evolve passively.  The passive evolution model
is a specific pure luminosity evolution model, where the evolution rate $Q$ should
follow that from an aging stellar population.  Arnouts et al. (2007) estimated
that the passive evolution for the $K$ band was $Q \sim 0.7$--1.0 from early to
late-type galaxies.  If we assume that the [3.6] and [4.5] bands have similar passive
evolution rates, our late and total LF evolution rates ($Q\sim0.9$--1.2) are 
consistent with the passive evolution model, while our early-type LF evolution rates 
($Q\sim1.3$--1.4) are $\sim1.5\sigma$ 
faster than the predictions.
If we include the data from Kochanek et al. (2001) and Arnouts et al. (2007),
the $Q$ value for early-type galaxies at [3.6] is slightly slower with $Q=1.20\pm0.16$.
We note that the $B$-band evolution rate for early-type galaxies, $Q\sim0.9$, from Brown et al. (2007)
is also slower than our estimate.

Our greatest concern in these estimates is that redshift-dependent biases in the total
magnitudes are mimicking evolution.  In our original calculation, we simply used the total
\verb+mag_auto+ magnitudes from the IRAC Shallow Survey and found still faster evolution
rates with $\Delta Q = +0.3$.  This drove our investigation of the difference in the 
aperture and \verb+mag_auto+ total magnitudes where we found a difference of about 0.15 mag
over the range from $z=0$ to 0.5 between the optical and mid-IR photometry.
That led us to the present approximation (see \S\ref{sec:sample}).
This needs to be investigated further, but a complete reanalysis of the survey photometry
is well beyond the scope of our analysis.  However, our investigations in Appendix A suggest that 
the redshift dependent magnitude definition problem in our current scheme is less severe, with 
systematic uncertainties of $\Delta Q < 0.1$.

\subsection{Density Evolution}

\begin{figure}
\epsscale{1}
\plotone{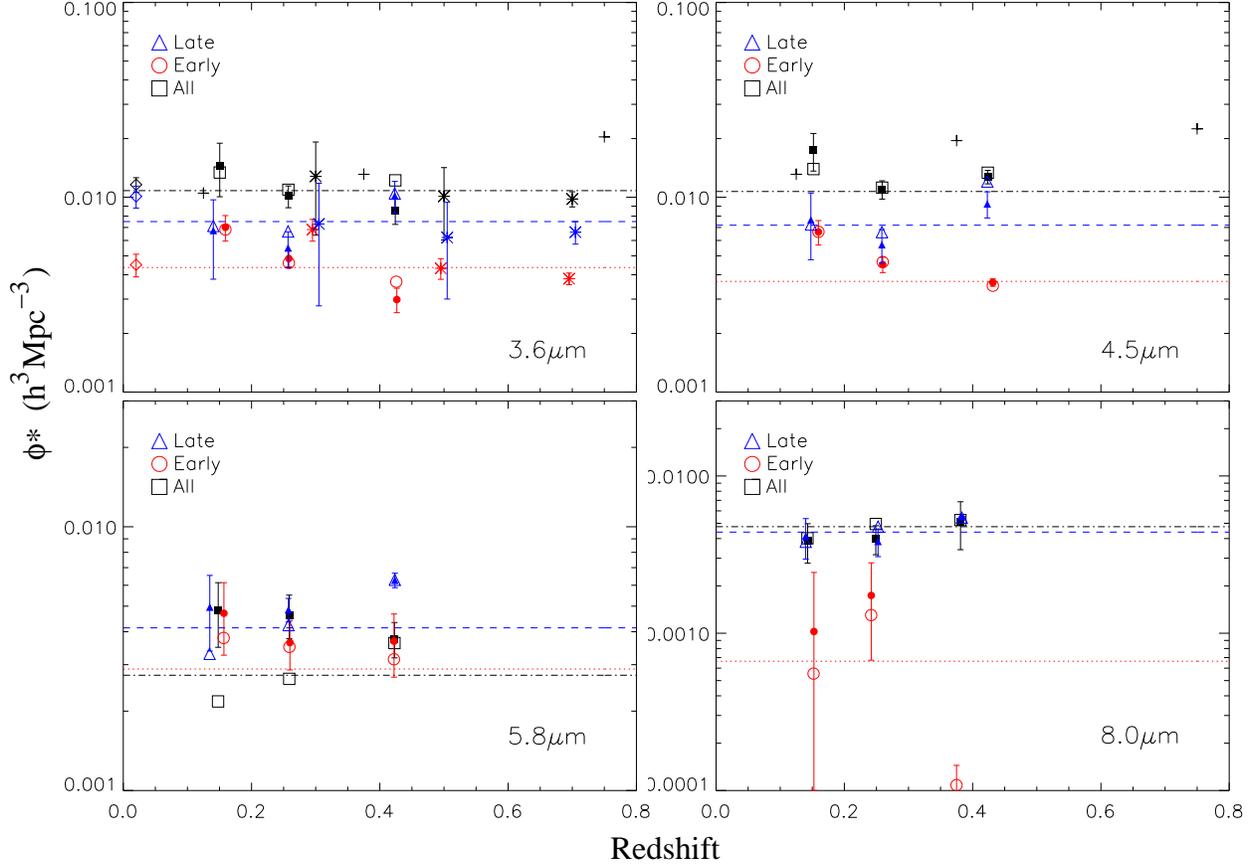}
\caption{The evolution of $\phi_*$ in the IRAC channels. The symbols and lines styles are the same as in Figure~\ref{fig:mstar}.  Babbedge et al. (2006) did not provide error bars for their estimates of $\phi_*$.
The cosmic variances of approximately 20\%, 15\%, and 10\% for the redshift bins of $z<0.2$, $0.2<z<0.35$, and $0.35<z<0.6$ are not included in the uncertainties.
\label{fig:nstar}}
\end{figure}

Figure~\ref{fig:nstar} shows the evolution of $\phi_*$ for the same three methods discussed in \S\ref{sec:method}.  In the global STY fit, $\phi_*$ is a constant.  In the STY fit to the individual
redshift bins combined with a Schechter function fit to the SWML LF, the $\phi_*$ are allowed to differ between
the redshift bins.
We estimated the cosmic variance in our sample using the estimator
\begin{equation}
\sigma_V^2 = \frac{1}{V^2}\int \xi(\left|{\bf{r}}_1-{\bf{r}}_2\right|)\, dV_1dV_2
\end{equation}
(e.g., Peebles 1980; Somerville et al. 2004).
We adopted a power-law correlation function $\xi = (r/r_0)^{-{\gamma}}$ with
$r_0 = 5.6, 5.7$, and 3.6 $h^{-1}$Mpc and $\gamma=1.8$, 2.1, and 1.7, for total, early, and late-type galaxies, respectively, measured from the SDSS survey (Zehavi et al. 2005).
In our three redshift bins
of $z<0.2$, $0.2<z<0.35$, $0.35<z<0.6$ and the total range $z<0.6$, we obtained cosmic variance estimates 
of 20\%, 15\%, 10\%, and 8\% for the total population, 18\%, 13\%, 8\%, and 7\% for the early-types, and 15\%, 11\%, 8\%, and 6\% for the late-type galaxies.
The cosmic variances we obtained are 
comparable to the statistical uncertainties in $\phi_*$ for the SWML LFs.

In general, we see no convincing evidence for density evolution, with the exception
of the early-type galaxies in the [3.6] and [4.5] bands, where we seem to see
a steady decline.  
For the STY method (fixed $\phi_*$) we obtained $\phi_* = 0.44\pm0.05$ and $0.37\pm0.04$ ($10^{-2}h^3{\rm Mpc}^{-3}$) 
for early-type galaxies in [3.6] and [4.5] for the full redshift range of $z\le0.6$.
Including the uncertainties from cosmic variance, 
the $\phi_*$ values from SWML for early-type galaxies are 
higher than the STY values by 1.3$\sigma$ and 1.6$\sigma$ in the first redshift bin $z<0.2$ 
for [3.6] and [4.5], and lower than the STY value by 2.0$\sigma$ in the third bin $0.35<z<0.6$ for [3.6].
For the second bin $0.2<z<0.35$, the two methods yielded consistent results.
However, this trend does not extend to $z=0$, when we compare to the $\phi_*$ found in the 
local 2MASS sample (Kochanek et al. 2001), and it suggests that the low redshift
point from AGES is high due to cosmic variance rather than due to rapid evolution. 
We note that the early-type galaxies in Kochanek et al. (2001) are morphologically
selected, which might also cause the difference.
The early-type sample in Arnouts et al. (2007) is also based on the SED fitting, and
there is a modest decline of $\phi_*$ with redshift in their three bins as well, 
consistent with our trend, 
but the absolute values of $\phi_*$ do not match between the two samples.
For the remaining cases there is no evidence for any significant density
evolution.  

We can also compare the measurements of density evolution to those in optical bands.
While the late-type galaxies are generally found to have no significant density 
evolution within $z<1$ consistent with our results, there are several studies suggesting
a significant density evolution for early-type galaxies within $z<1$ from 40\% to 400\%
(Zucca et al. 2006; Faber et al. 2007; Bell et al. 2004).  However, there is also
evidence for little density evolution in early-type galaxies (Brown et al. 2007).
Our AGES LFs at [3.6] suggest possible density evolution for early-type galaxies at $z < 0.6$;
however, adding the data points from Arnouts et al. (2007) and Kochanek et al. (2001)
and considering the cosmic variance, the combined data do not provide a definitive answer.

\subsection{The Luminosity Density and Its Evolution}

\begin{figure}
\epsscale{1}
\plotone{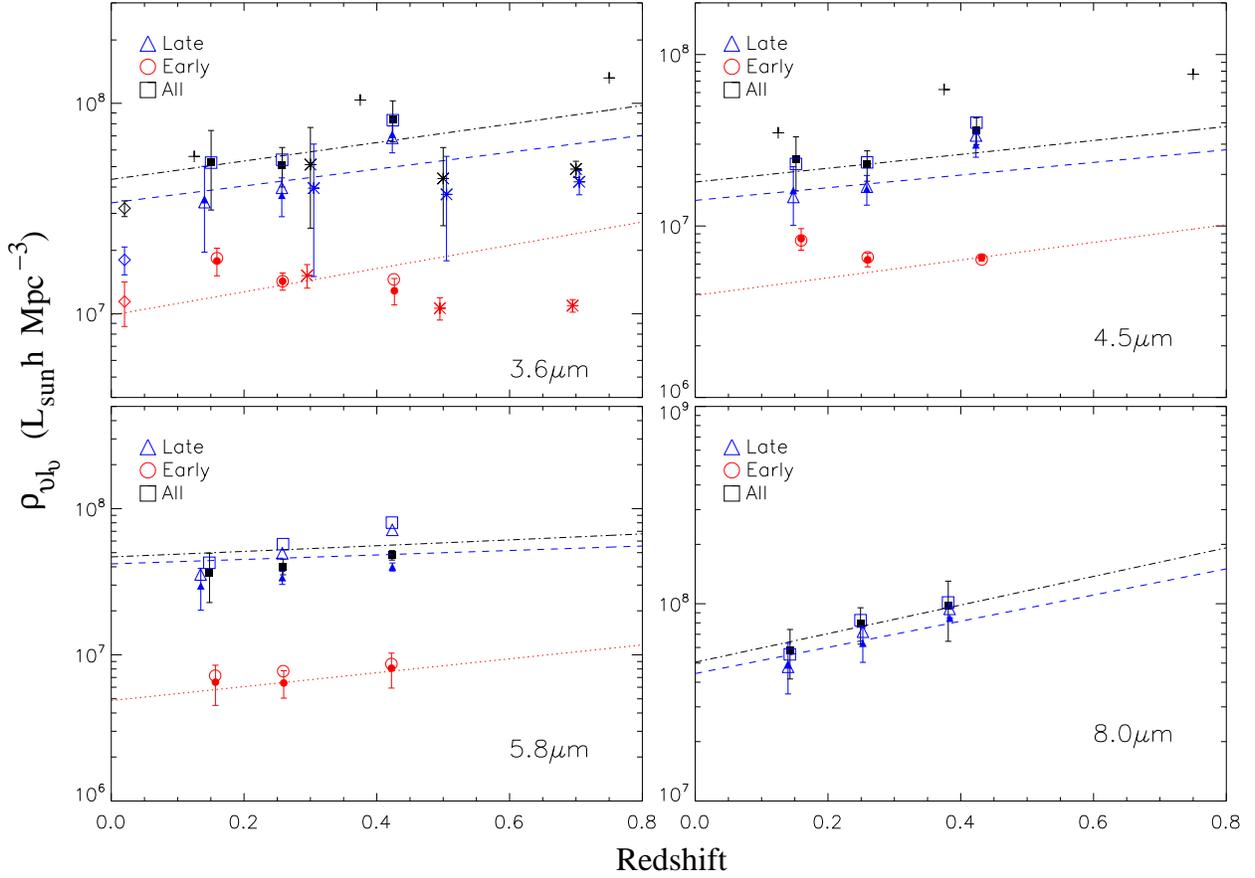}
\caption{The evolution of the luminosity density in the IRAC channels.  
The symbols and line styles are the same as in Figure~\ref{fig:mstar}.  
At [3.6] and [4.5], the late-type and total luminosity density evolution is
consistent with a passive evolution model, while the early-type galaxies
show deviations from the model.
Fitting all the data for the early-type galaxies in [3.6], we found that the stellar mass has 
increased by $87\pm30$\% from $z=0.7$ to $z=0$ compared to the passive evolution 
model.
\label{fig:lstar}}
\end{figure}

Assuming that we can extrapolate the luminosity functions beyond the magnitude
limits of the samples, we can compute the luminosity densities from the LFs
as $\rho_{\nu L_{\nu}} = \phi_* (\nu L_{\nu, *}) \Gamma(2+\alpha)$, where we
obtain $L_{\nu, *}$ through our $M_*$ and IRAC zero points.  
The results will be insensitive
to the value of the faint end slope when $\alpha \sim -1$.  However, when the 
faint end slope is steep, the uncertainties introduced by $\alpha$ are large. 
In particular, for the case of the early-type galaxies
at [8.0], the result is divergent because $\alpha \simeq -2.0$.  Fig.~\ref{fig:lstar}
shows the luminosity densities for our three standard methods, except for the
early-type galaxies at [8.0] because of the large uncertainties.  
In general, we see a trend of increasing
luminosity density with higher redshifts, with the weakest trends for the
early-type galaxies at [3.6] and [4.5]. 

Even passive evolution models predict that the luminosity density increases with redshift.
Since the evolution of $M_*$ and $\phi_*$ for late-type and total galaxies in [3.6] and [4.5] are 
consistent with the predictions from passive evolution (constant $\phi_*$ and $Q\sim1$), 
the evolution in luminosity
density must also be consistent with the predictions.
For early-type galaxies, the evolution of luminosity density will tend to provide a more 
robust test for the passive evolution models than examinations of the
individual Schechter parameters because they are less sensitive to the strong correlations
between $\alpha$ and $M_*$.  
 Combining the data from Kochanek et al. (2001) and
Arnouts et al. (2007), the luminosity density evolution in [3.6] is at least flat. 
We compare this with the expected luminosity density evolution from the passive evolution model.
In the passive evolution model, $\alpha$ and $\phi_*$ are constants, and hence the
luminosity density scales with $L_*$.  Using the $K$-band passive evolution rate of $Q=0.7$ 
for early-type galaxies (Arnouts et al. 2007), $L_*$ should dim by 0.5~mag from $z=0.7$ to 0.
The constant trend of the [3.6] band luminosity density
deviates from the passive evolution model and indicate an increase of stellar
mass of $40\pm20$\% from $z=0.7$ to $z=0$ assuming a constant mass-to-light ratio at $z=0$,
when we compare the first and last data points from Kochanek et al. (2001) and Arnouts et al. (2007).  
We fit all the data with a power-law and obtained 
\begin{equation}
\log \frac{\rho_{\nu L_{\nu}}}{L_{\odot} h {\rm Mpc}^{-3}} = (7.18\pm0.04) - (0.20\pm0.07) z.
\end{equation}
This indicates an increase of stellar mass of $87\pm30$\% from $z=0.7$ to 0 for early-type
galaxies.
If we remove the 2MASS point based on the morphological
definition and use only the 6 remaining points based on a SED definition for early-type galaxies,
the luminosity density evolution shows still larger differences from a passive evolution model. 
Our results are consistent with the flat $B$-band luminosity density evolution for early-type
galaxies (Bell et al. 2004; Faber et al. 2007; Brown et al. 2007), and the $K$-band analysis
of Arnouts et al. (2007) who find the stellar mass for early-type galaxies has increased by 
100\% from $z=1.2$ to 0.

\begin{figure}
\epsscale{1}
\plotone{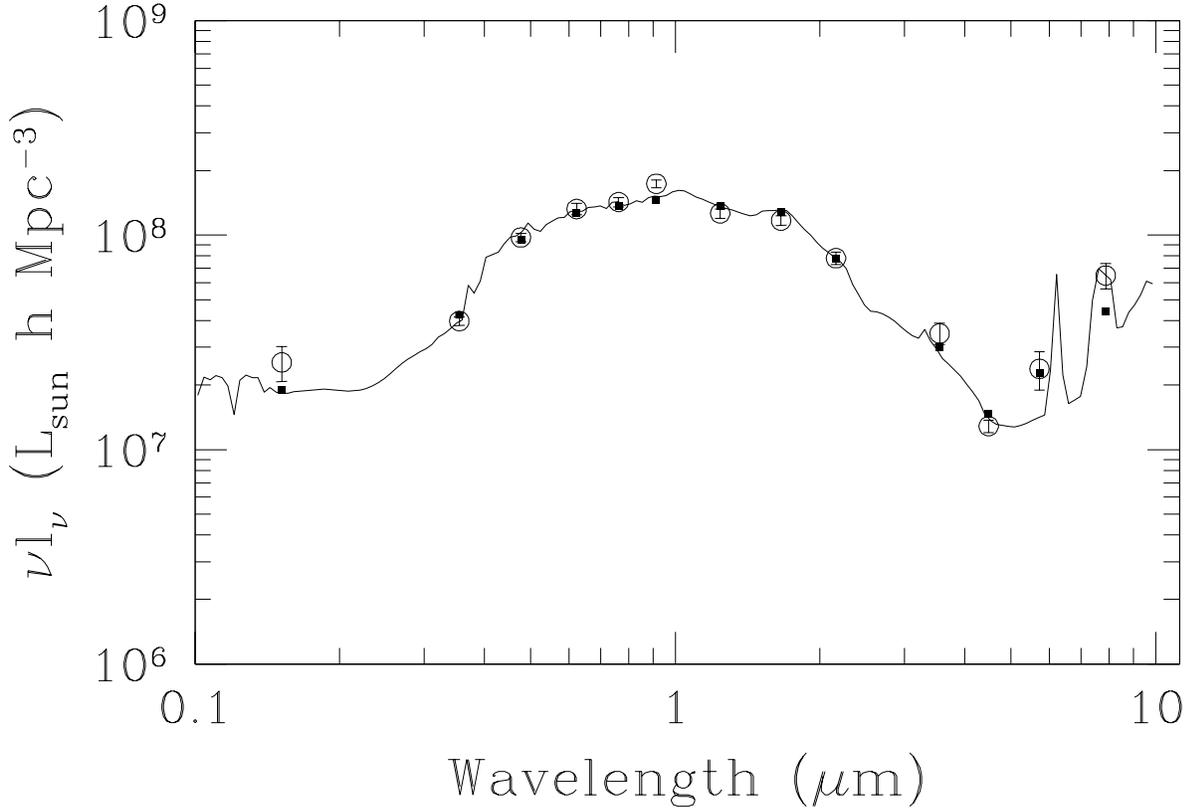}
\caption{The galaxy luminosity density (circles) as a function of wavelength at $z \simeq 0.1$ from the UV to the mid-IR, using the local LFs 
determined from GALEX (Arnouts et al.\ 2005), SDSS (Bell et al.\ 2003), 2MASS (Jones et al.\ 2006), and our LFs for the Spitzer/IRAC bands.
The solid line is the best-fit spectrum, and the squares are the convolution of the best-fit SED and the different filter bandpasses.
This excludes any contribution from known AGN.  The 
luminosity density spectrum is that of a mildly star forming galaxy (with an early-type fraction of $\hat{e}=0.25$), where the emission drops from
the near-IR to a minimum near 5$\mu$m and rises again due to PAH emission. 
\label{fig:lden}}
\end{figure}

In Fig.~\ref{fig:lden} we shift the luminosity densities to $z=0.1$ and combine them 
with earlier results scaled to that redshift at shorter wavelengths from the near-IR through the UV,
based on results from GALEX, SDSS and 2MASS (Arnouts et al.\ 2005, Bell et al.\ 2003, 
Jones et al.\ 2006).   The ``spectrum'' given by the luminosity density is typical
of a moderately star forming galaxy, as we illustrate by fitting the implied SED
with the template models from Assef et al. (2008).  The luminosity density spectrum
has an early-type fraction of $\hat{e}=0.25$. 

\subsection{Comparison with a Semi-Analytical Model}
We can also compare our LFs with the predictions from the recent semi-analytical models of
 Lacey et al. (2008).
Lacey et al.\ (2008) combined a semi-analytical hierarchical galaxy formation model based on CDM, 
a theoretical stellar population synthesis model for stellar emission, and a theoretical radiative transfer
model for dust absorption and emission.  They also assumed a top-heavy IMF in star-bursts and a normal
solar neighborhood IMF for quiescent star formation.  The model was tuned to reproduce the $B$, $K$, and 60\um\ LFs,
and several observed interrelationships between galaxy luminosity, gas mass, metallicity, size, and the fraction of
spheroidal galaxies.

\begin{figure}
\epsscale{1}
\plotone{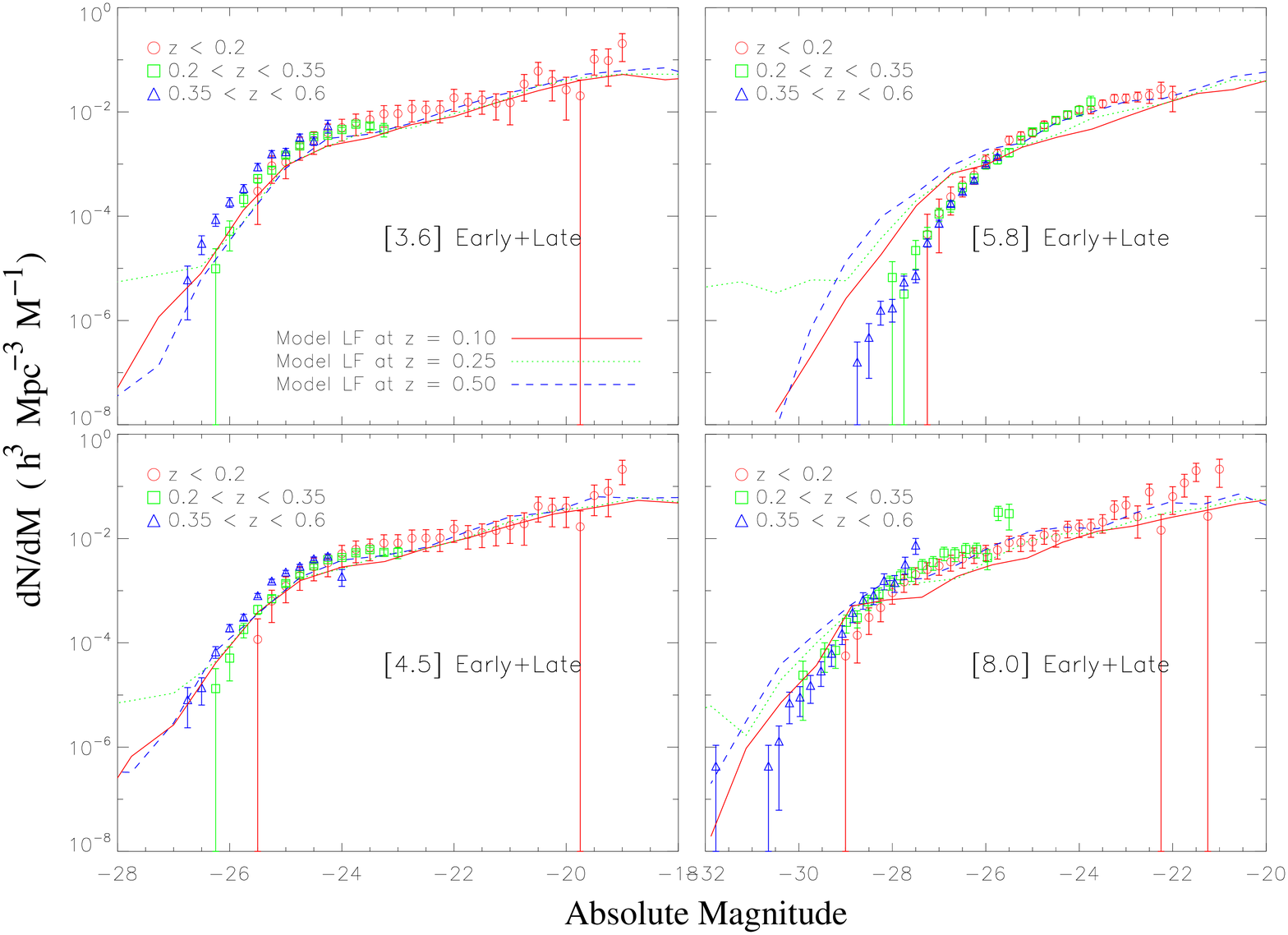}
\caption{Comparisons between the [3.6]--[8.0] LFs from AGES at $z<0.25$, $0.2<z<0.35$, and $0.35<z<0.6$ and 
the theoretical models from Lacey et al. (2008) at $z=0.1$, 0.25, and 0.5.  The match between the models and the observations is best at [3.6] and [4.5], and it is worst at [5.8].
\label{fig:lacey}}
\end{figure}

Figure~\ref{fig:lacey} shows the comparison between our SWML LFs at $z<0.2$, $0.2<z<0.35$, and $0.35<z<0.6$ 
to the theoretical models at $z=0.1$, $z=0.25$, and $z=0.5$.  
Note that there are small mismatches
between our median redshifts of $z=0.15$, 0.25, and 0.45 and those of the models.
In general, the models match our observed [3.6] and [4.5] LFs well, they are roughly consistent at [8.0], 
and they fail to reproduce the [5.8] LFs.  At [3.6] and [4.5], the shape of model LFs is consistent with the Schechter
form at $M \simlt -22$~mag, and steepen at the faint end consistent with the tail of our SWML LFs.  
The model slightly over-predicts
the [8.0] LFs at the bright end, and under-predicts the faint end LFs.  
The significantly worse match at [5.8]
is likely due to problems with the PAH features in the theoretical models.
This affects the [5.8] more than the other bands because at [5.8] the sample 
is composed of a mixture of stellar and ISM emission 
and has the largest scatter in its K-corrections (see Figure~\ref{fig:kcor}).

We can also compare the LF evolution between the models and our measurements.
At [3.6] and [4.5], the model predicts little LF evolution at the bright end,
although the match of the evolution rates at [8.0] is better.
Lacey et al. (2008) also modeled the early and late-type LFs separately, 
where the late-type galaxy LFs are similar to the total LFs except for the normalization,
and the early-type LFs have significant flatter faint end slopes.  This is also consistent with our measurements.

\begin{deluxetable}{cccccccc}
\tabletypesize{\scriptsize}
\tablecolumns{8}
\tablewidth{0pt}
\tablecaption{STY Parametric Fitting Results for the Mid-Infrared Luminosity Functions at $z<0.6$ \label{tab:styfit}}
\tablehead{
\colhead{Band} &
\colhead{Type} &
\colhead{N Galaxies} &
\colhead{Median $z$} &
\colhead{$\alpha$} &
\colhead{$M_*-5\log h$} & 
\colhead{$Q$} &
\colhead{$\phi_*$} \\
\colhead{} &
\colhead{} &
\colhead{} &
\colhead{} &
\colhead{} &
\colhead{ (mag, at $z=0.25$)} &
\colhead{} &
\colhead{($10^{-2}h^3$Mpc$^{-3}$)} 
}
\startdata
$[3.6]$ & all   & 4905 & 0.235 & $-1.12\pm0.16$ & $-24.29\pm0.18$ & $1.2\pm0.4$ & $1.08\pm0.03$ \\
$[3.6]$ & early & 2222 & 0.253 & $-0.63\pm0.29$ & $-24.18\pm0.22$ & $1.4\pm0.5$ & $0.44\pm0.05$ \\
$[3.6]$ & late  & 2683 & 0.216 & $-1.40\pm0.18$ & $-24.27\pm0.20$ & $1.0\pm0.7$ & $0.75\pm0.02$ \\
$[4.5]$ & all   & 5847 & 0.246 & $-1.02\pm0.13$ & $-24.20\pm0.16$ & $1.1\pm0.4$ & $1.07\pm0.03$ \\
$[4.5]$ & early & 2422 & 0.261 & $-0.60\pm0.28$ & $-24.08\pm0.25$ & $1.3\pm0.5$ & $0.37\pm0.04$ \\
$[4.5]$ & late  & 3425 & 0.242 & $-1.29\pm0.17$ & $-24.26\pm0.21$ & $0.9\pm0.7$ & $0.72\pm0.02$ \\
$[5.8]$ & all   & 4367 & 0.239 & $-1.85\pm0.13$ & $-26.03\pm0.17$ & $-0.3\pm0.6$\phs & $0.27\pm0.01$ \\
$[5.8]$ & early & 1741 & 0.253 & $-1.33\pm0.28$ & $-24.86\pm0.27$ & $1.2\pm0.8$ & $0.29\pm0.03$ \\
$[5.8]$ & late  & 2626 & 0.220 & $-1.63\pm0.15$ & $-25.96\pm0.20$ & $0.4\pm0.7$ & $0.41\pm0.01$ \\
$[8.0]$ & all   & 3802 & 0.195 & $-1.42\pm0.12$ & $-27.84\pm0.26$ & $1.8\pm1.1$ & $0.48\pm0.01$ \\
$[8.0]$ & early & 494  & 0.191 & $-2.03\pm0.47$ & $-26.86\pm0.73$ & $1.8\pm3.5$ & $0.07\pm0.01$ \\
$[8.0]$ & late  & 3308 & 0.197 & $-1.35\pm0.09$ & $-27.86\pm0.17$ & $1.7\pm0.8$ & $0.44\pm0.01$ \\
\enddata

\tablecomments{The galaxy samples are fit with the STY method and a pure luminosity evolution model with $\Delta M_* = Qz$.  The cosmic variance in the redshift range of $z<0.6$ is 8\%, which is
not included in the statistical uncertainties given for $\phi_*$ in this table.}
\end{deluxetable}

\begin{deluxetable}{cccccccccc}
\rotate
\tabletypesize{\scriptsize}
\tablecolumns{10}
\tablewidth{0pt}
\tablecaption{Schechter Function Fits for Binned SWML LFs \label{tab:swmlfit}}
\tablehead{
\colhead{} &
\colhead{} &
\colhead{} &
\multicolumn{2}{c}{$z \le 0.2$} &
\multicolumn{2}{c}{$0.2 \le z \le 0.35$} &
\multicolumn{2}{c}{$0.35 \le z \le 0.6$} &
\colhead{} \\
\cline{4-5}
\cline{6-7}
\cline{8-9}
\colhead{Band} &
\colhead{Type} &
\colhead{$\alpha$} &
\colhead{$M_*-5\log h$} &
\colhead{$\phi_*$} &
\colhead{$M_*-5\log h$} &
\colhead{$\phi_*$} &
\colhead{$M_*-5\log h$} &
\colhead{$\phi_*$} &
\colhead{$\chi^2(dof)$} \\
\colhead{} &
\colhead{} &
\colhead{} &
\colhead{(mag)} &
\colhead{($10^{-2}h^3$Mpc$^{-3}$)} &
\colhead{(mag)} &
\colhead{($10^{-2}h^3$Mpc$^{-3}$)} &
\colhead{(mag)} &
\colhead{($10^{-2}h^3$Mpc$^{-3}$)} &
\colhead{} 
}
\startdata
$[3.6]$ & all   & $-1.12\pm0.13$ & $-24.09\pm0.24$ & $1.45\pm0.44$ & $-24.34\pm0.05$ & $1.01\pm0.13$ & $-24.63\pm0.09$ & $0.85\pm0.13$ & 32.6(44) \\
$[3.6]$ & early & $-0.57\pm0.14$ & $-23.97\pm0.19$ & $0.70\pm0.11$ & $-24.13\pm0.11$ & $0.48\pm0.05$ & $-24.54\pm0.08$ & $0.30\pm0.04$ & 28.8(36) \\
$[3.6]$ & late  & $-1.42\pm0.14$ & $-24.13\pm0.19$ & $0.67\pm0.30$ & $-24.41\pm0.09$ & $0.55\pm0.11$ & $-24.46\pm0.09$ & $1.02\pm0.19$ & 28.7(42) \\
$[4.5]$ & all   & $-0.97\pm0.11$ & $-23.86\pm0.18$ & $1.74\pm0.38$ & $-24.23\pm0.06$ & $1.10\pm0.12$ & $-24.42\pm0.04$ & $1.27\pm0.10$ & 29.9(45) \\
$[4.5]$ & early & $-0.57\pm0.13$ & $-23.96\pm0.20$ & $0.66\pm0.10$ & $-24.07\pm0.10$ & $0.45\pm0.04$ & $-24.33\pm0.03$ & $0.36\pm0.02$ & 36.4(37) \\
$[4.5]$ & late  & $-1.34\pm0.14$ & $-24.04\pm0.28$ & $0.76\pm0.29$ & $-24.38\pm0.13$ & $0.57\pm0.11$ & $-24.50\pm0.07$ & $0.92\pm0.14$ & 30.6(46) \\
$[5.8]$ & all   & $-1.60\pm0.07$ & $-25.79\pm0.19$ & $0.48\pm0.13$ & $-25.70\pm0.08$ & $0.46\pm0.09$ & $-25.76\pm0.06$ & $0.38\pm0.06$ & 36.2(46) \\
$[5.8]$ & early & $-1.17\pm0.18$ & $-24.53\pm0.24$ & $0.47\pm0.14$ & $-24.79\pm0.07$ & $0.36\pm0.08$ & $-25.03\pm0.04$ & $0.37\pm0.10$ & 12.6(36) \\
$[5.8]$ & late  & $-1.46\pm0.11$ & $-25.71\pm0.15$ & $0.50\pm0.16$ & $-25.88\pm0.13$ & $0.49\pm0.05$ & $-25.78\pm0.08$ & $0.63\pm0.04$ & 20.9(44) \\
$[8.0]$ & all   & $-1.46\pm0.07$ & $-27.69\pm0.08$ & $0.39\pm0.11$ & $-28.00\pm0.13$ & $0.40\pm0.08$ & $-27.95\pm0.13$ & $0.51\pm0.17$ & 42.1(61) \\
$[8.0]$ & early & $-1.85\pm0.21$ & $-26.50\pm0.76$ & $0.10\pm0.14$ & $-26.37\pm0.29$ & $0.17\pm0.11$ & $-27.29\pm1.15$ & $0.01\pm0.01$ & 13.5(32) \\
$[8.0]$ & late  & $-1.34\pm0.09$ & $-27.62\pm0.14$ & $0.42\pm0.12$ & $-27.99\pm0.05$ & $0.38\pm0.08$ & $-27.91\pm0.26$ & $0.55\pm0.01$ & 36.8(59) \\
\enddata

\tablecomments{In each sample, the SWML LFs in the three redshift bins are jointly fit with 
Schechter functions, 
where we fixed the faint end slope $\alpha$ to be the same in all bins but allowed the 
$M_*$ and the normalization $\phi_*$ to differ.
The cosmic variances are 20\%, 15\%, and 10\% for the redshift bins of $z<0.2$, $0.2<z<0.35$, and $0.35<z<0.6$, which are not included in the error-bars of $\phi_*$ in this table.}
\end{deluxetable}

\begin{deluxetable}{cccccc}
\tabletypesize{\scriptsize}
\tablecolumns{6}
\tablewidth{0pt}
\tablecaption{Binned SWML LFs \label{tab:swml}}
\tablehead{
\colhead{Band} &
\colhead{Type} &
\colhead{Mag}  &
\multicolumn{3}{c}{$dN/dM$ ($h^3$Mpc$^{-3}$mag$^{-1}$)} \\
\cline{4-6}
\colhead{} &
\colhead{} &
\colhead{} &
\colhead{$z \le 0.2$} &
\colhead{$0.2 \le z \le 0.35$} &
\colhead{$0.35 \le z \le 0.6$} 
}
\startdata
$[3.6]$ & All & $-19.00$ & 2.0E$-$1 (1.1E$-$1) & \nodata & \nodata \\
$[3.6]$ & All & $-19.25$ & 9.6E$-$2 (6.5E$-$2) & \nodata & \nodata \\
$[3.6]$ & All & $-19.50$ & 1.0E$-$1 (5.2E$-$2) & \nodata & \nodata \\
$[3.6]$ & All & $-19.75$ & 2.0E$-$2 (2.1E$-$2) & \nodata & \nodata \\
$[3.6]$ & All & $-20.00$ & 2.7E$-$2 (2.0E$-$2) & \nodata & \nodata \\
$[3.6]$ & All & $-20.25$ & 3.9E$-$2 (2.3E$-$2) & \nodata & \nodata \\
$[3.6]$ & All & $-20.50$ & 6.1E$-$2 (2.8E$-$2) & \nodata & \nodata \\
$[3.6]$ & All & $-20.75$ & 3.4E$-$2 (1.8E$-$2) & \nodata & \nodata \\
$[3.6]$ & All & $-21.00$ & 1.5E$-$2 (9.4E$-$3) & \nodata & \nodata \\
$[3.6]$ & All & $-21.25$ & 1.5E$-$2 (7.6E$-$3) & \nodata & \nodata \\
$[3.6]$ & All & $-21.50$ & 1.7E$-$2 (8.0E$-$3) & \nodata & \nodata \\
$[3.6]$ & All & $-21.75$ & 1.5E$-$2 (7.1E$-$3) & \nodata & \nodata \\
$[3.6]$ & All & $-22.00$ & 1.9E$-$2 (8.5E$-$3) & \nodata & \nodata \\
$[3.6]$ & All & $-22.25$ & 1.1E$-$2 (5.1E$-$3) & \nodata & \nodata \\
$[3.6]$ & All & $-22.50$ & 1.1E$-$2 (5.0E$-$3) & \nodata & \nodata \\
$[3.6]$ & All & $-22.75$ & 1.1E$-$2 (5.1E$-$3) & \nodata & \nodata \\
$[3.6]$ & All & $-23.00$ & 9.3E$-$3 (4.1E$-$3) & \nodata & \nodata \\
$[3.6]$ & All & $-23.25$ & 9.2E$-$3 (4.0E$-$3) & 4.7E$-$3 (1.4E$-$3) & \nodata \\
$[3.6]$ & All & $-23.50$ & 7.1E$-$3 (3.2E$-$3) & 5.4E$-$3 (7.3E$-$4) & \nodata \\
$[3.6]$ & All & $-23.75$ & 6.3E$-$3 (2.8E$-$3) & 5.8E$-$3 (5.8E$-$4) & \nodata \\
$[3.6]$ & All & $-24.00$ & 5.0E$-$3 (2.2E$-$3) & 4.6E$-$3 (4.5E$-$4) & \nodata \\
$[3.6]$ & All & $-24.25$ & 4.0E$-$3 (1.8E$-$3) & 3.6E$-$3 (3.4E$-$4) & 5.4E$-$3 (1.5E$-$3) \\
$[3.6]$ & All & $-24.50$ & 2.8E$-$3 (1.3E$-$3) & 3.4E$-$3 (3.1E$-$4) & 2.8E$-$3 (5.1E$-$4) \\
$[3.6]$ & All & $-24.75$ & 2.2E$-$3 (1.0E$-$3) & 2.2E$-$3 (2.3E$-$4) & 3.3E$-$3 (5.0E$-$4) \\
$[3.6]$ & All & $-25.00$ & 1.1E$-$3 (5.6E$-$4) & 1.5E$-$3 (1.8E$-$4) & 1.7E$-$3 (2.7E$-$4) \\
$[3.6]$ & All & $-25.25$ & 9.3E$-$4 (5.0E$-$4) & 7.6E$-$4 (1.2E$-$4) & 1.6E$-$3 (2.5E$-$4) \\
$[3.6]$ & All & $-25.50$ & 3.0E$-$4 (2.3E$-$4) & 5.3E$-$4 (9.8E$-$5) & 8.9E$-$4 (1.4E$-$4) \\
$[3.6]$ & All & $-25.75$ & \nodata & 2.1E$-$4 (6.0E$-$5) & 3.4E$-$4 (6.3E$-$5) \\
$[3.6]$ & All & $-26.00$ & \nodata & 5.1E$-$5 (3.0E$-$5) & 1.9E$-$4 (3.9E$-$5) \\
$[3.6]$ & All & $-26.25$ & \nodata & 9.8E$-$6 (1.4E$-$5) & 8.7E$-$5 (2.3E$-$5) \\
$[3.6]$ & All & $-26.50$ & \nodata & \nodata & 3.0E$-$5 (1.2E$-$5) \\
$[3.6]$ & All & $-26.75$ & \nodata & \nodata & 6.0E$-$6 (5.0E$-$6) \\
\enddata
\tablecomments{Table~\ref{tab:swml} is in its entirety in the electronic edition of the journal.
A portion is shown here for guidance regarding its form and content.}
\end{deluxetable}

\begin{deluxetable}{lcccccc}
\tabletypesize{\scriptsize}
\tablecolumns{7}
\tablewidth{0pt}
\tablecaption{Comparison with Other IR LFs \label{tab:comp}}
\tablehead{
\colhead{Sample} &
\colhead{Band} &
\colhead{Type} &
\colhead{redshift} &
\colhead{$\alpha$} &
\colhead{$M_*-5\log h$} &
\colhead{$\phi_*$} \\
\colhead{} &
\colhead{} &
\colhead{} &
\colhead{} &
\colhead{} &
\colhead{(mag)} &
\colhead{($10^{-2}h^3$Mpc$^{-3}$)} 
}
\startdata
Kochanek et al. (2001) & converted to [3.6] & all   & 0.02 & $-1.09\pm0.06$ & $-23.85\pm0.05$ & $1.16\pm0.10$ \\
                       &                    & early &      & $-0.92\pm0.10$ & $-23.88\pm0.10$ & $0.45\pm0.06$ \\
                       &                    & late  &      & $-0.87\pm0.09$ & $-23.52\pm0.09$ & $1.01\pm0.13$ \\
Arnouts et al. (2007)  & converted to [3.6] & all   & 0.3  & $-1.1\pm0.2$   & $-24.3\pm0.6$  & $1.3\pm0.6$ \\
                       &                    &       & 0.5  & $-1.1\pm0.2$   & $-24.4\pm0.3$  & $1.0\pm0.4$ \\
                       &                    &       & 0.7  & $-1.12\pm0.06$ & $-24.5\pm0.1$  & $1.0\pm0.1$ \\
                       &                    & early & 0.3  & $-0.3\pm0.2$   & $-23.8\pm0.3$  & $0.68\pm0.09$ \\
                       &                    &       & 0.5  & $-0.3\pm0.2$   & $-23.9\pm0.2$  & $0.43\pm0.05$ \\
                       &                    &       & 0.7  & $-0.29\pm0.12$ & $-24.1\pm0.1$  & $0.38\pm0.03$ \\
                       &                    & late  & 0.3  & $-1.3\pm0.2$   & $-24.4\pm0.6$  & $0.7\pm0.5$ \\
                       &                    &       & 0.5  & $-1.3\pm0.2$   & $-24.5\pm0.3$  & $0.6\pm0.3$ \\
                       &                    &       & 0.7  & $-1.30\pm0.07$ & $-24.6\pm0.1$  & $0.66\pm0.09$ \\
Babbedge et al. (2006) & [3.6]           & all & 0.00--0.25 & $-0.9$         & $-23.9\pm0.1$ & $1.0$ \\
                       &                 &     & 0.25--0.50 & $-1.0$         & $-24.3\pm0.1$ & $1.3$ \\
                       &                 &     & 0.50--1.00 & $-0.9$         & $-24.1\pm0.1$ & $2.0$ \\
                       & [4.5]           &     & 0.00--0.25 & $-0.9$         & $-23.9\pm0.1$ & $1.3$ \\
                       &                 &     & 0.25--0.50 & $-1.0$         & $-24.1\pm0.1$ & $2.0$ \\
                       &                 &     & 0.50--1.00 & $-1.1$         & $-24.1\pm0.1$ & $2.2$ \\
Huang et al. (2007)    & [8.0]           & all &  0.0--0.3  & $-1.38\pm0.04$ & $-27.69\pm0.08$ & $0.32$ \\
                       &                 & PAH &            & $-1.26\pm0.04$ & $-27.57\pm0.08$ & $0.35$ \\
This Paper             & [3.6]           & all & 0.24       & $-1.12\pm0.16$ & $-24.29\pm0.18$ & $1.08\pm0.03$ \\
                       & [4.5]           &     & 0.25       & $-1.02\pm0.13$ & $-24.20\pm0.16$ & $1.07\pm0.03$ \\
                       & [5.8]           &     & 0.24       & $-1.85\pm0.13$ & $-26.03\pm0.17$ & $0.27\pm0.01$ \\
                       & [8.0]           &     & 0.20       & $-1.42\pm0.12$ & $-27.84\pm0.26$ & $0.48\pm0.01$ \\

\enddata
\end{deluxetable}

\section{Discussion}

The spectroscopy from the AGES survey has allowed us to measure the mid-infrared (3.6, 4.5, 5.6, and 8.0$\mu$m)
luminosity functions for $z<0.6$ with greater precision than existing mid-IR surveys which have largely relied
on photometric redshifts.  The bluest bands agree well with local $K$-band 
luminosity functions, possibly with some effects from the PAH feature in the 3.6$\mu$m band.
The early and late-type
galaxies having similar characteristic magnitudes and the early-type galaxies have shallower faint end slopes. As
we move to the redder bands, the early-type galaxies exhibit fainter break magnitudes and steeper faint end 
slopes relative to the late-type galaxies.  In general our results agree well with other recent mid-IR studies
based largely on photometric redshifts by Franceschini et al. (2006), Babbedge et al. (2006) and Arnouts
et al. (2007).  Although we have better statistics and use only spectroscopic redshifts, our study
is limited to lower redshifts.  
The one major exception is that we find that Schechter function fits 
work reasonably well at 5.8 and 8.0$\mu$m, as Huang et al. (2007) also found for 8.0$\mu$m based on a sub-sample of galaxies with  $z<0.3$ in our field.  This is in disagreement with the power-law fits adapted by Babbedge et al. (2006)
for these bands.  

Photometric redshifts are known to work well for the typical galaxy (e.g., \S3.1, Babbedge et al.\ 2006)
which is why our luminosity functions broadly agree with those based
solely or largely on them.  Where spectroscopic redshifts have an
edge is on the wings of the luminosity function, where magnitude limited
samples have few objects because the high luminosity objects are rare
and the volume in which low luminosity objects can be found is small.
These parts of the luminosity function, well away from $L_*$, are
quickly altered given even a small number of catastrophic photo-z
redshift errors for the far more numerous $L_*$ galaxies.  The general
tendency will be to weaken the exponential cutoff of a Schechter
function at high luminosity and to steepen the slope at low luminosity.
While we have no direct evidence that this is the explanation, this
is exactly the trend of our differences with the longer wavelength
SWIRE luminosity functions (Babbedge et al.\ 2006).

We find no convincing evidence for density evolution in our sample, and a pure luminosity evolution model
appears to work reasonably well.
The $M_*$ for the total galaxy population evolves as $ Q  z$ with $Q \simeq 1.1$--1.2 in the [3.6] and [4.5] bands, 
which probe the stellar mass, and the evolution rates are consistent with the $K$-band passive evolution models of Arnouts et al. (2007, $Q\simeq1.0$).
We measured the evolution rate $Q \simeq 1.8$ in the [8.0] band, which is
sensitive to star formation and consistent with other estimates for the evolution of star formation (Hopkins 2004; Villar et al. 2008).
The rate of evolution agrees well with the scalings from 2MASS, Arnouts et al. (2007) and Babbedge et al. (2006) at $3.6\mu$m.
At [3.6] and [4.5], the evolution of $M_*$ and $\phi_*$ for early-type galaxies suggests possible 
deviations from passive evolution models, however, with large uncertainties.  
The evolution of luminosity density for early-type galaxies provides a more robust test for
deviations from the passive evolution model and suggests that the stellar mass for early-type
galaxies has increased by $87\pm30$\% from $z=0.7$ to $z=0$.
We also compared our LFs with the recent semi-analytical model from Lacey et al. (2008).
While the match between the model and observations is excellent at [3.6] and [4.5], 
it is worse at [8.0], and at [5.8] the model failed to reproduce the [5.8] LFs. 
We can also extend measurements of the galaxy luminosity density at $z=0.1$ into the mid-IR.  The 
luminosity density spectrum is that of a mildly star forming galaxy where the emission drops from
the near-IR to a minimum near 5$\mu$m and rises again due to PAH emission. 

Our results at low redshift would be significantly improved by combining our sample with a complete
redshift survey of the brighter mid-IR sources in the wider area SWIRE fields, to better constrain
the low redshift, high luminosity sources, and with a fainter sample in a narrow field (e.g. the
DEEP2 results for the Extended Groth Strip) to better constrain the faint end of the luminosity
function and extend the results to higher redshifts.  Within the Bo\"otes field itself we can
achieve many of the same goals using photometric redshifts. In particular, Assef et al. (2008)
have developed a set of templates that extend through all four IRAC bands, which would probably
give better results than most existing studies which have truncated their templates near
4.5$\mu$m band due to a lack of good templates for the longer wavelengths.  
Since our present analysis used the rest-frame 8$\mu$m results from the Assef et al. (2008) 
templates, which is based on data which extends only to 5$\mu$m for sources at 
$z=0.6$, it would also be
useful to extend the templates through the MIPS 24$\mu$m band.

\acknowledgements
We thank C. Lacey for providing the theoretical models for the IRAC band luminosity functions,
S. Arnouts and T. Babbedge for providing more detailed information on their results, and S. Willner for helpful discussions.
The AGES observations were obtained at the MMT Observatory, a joint facility of the Smithsonian
Institution and the University of Arizona.
We are grateful to the expert assistance of the staff of Kitt Peak National Observatory where the Bo\"otes field observations of the NDWFS were
obtained. The authors thank NOAO for supporting the NOAO Deep Wide-Field Survey. The research activities of A.~D. and B.~T.~J. are supported by NOAO, which
is operated by the Association of Universities for Research in Astronomy (AURA) under a cooperative agreement with the National Science Foundation. 
This work is based on observations made with the \emph{Spitzer Space Telescope}, which is operated by the
Jet Propulsion Laboratory, California Institute of Technology under a contract with NASA.
\clearpage

\appendix
\section{Standardizing the IRAC Photometry}
\begin{figure}
\epsscale{0.68}
\plotone{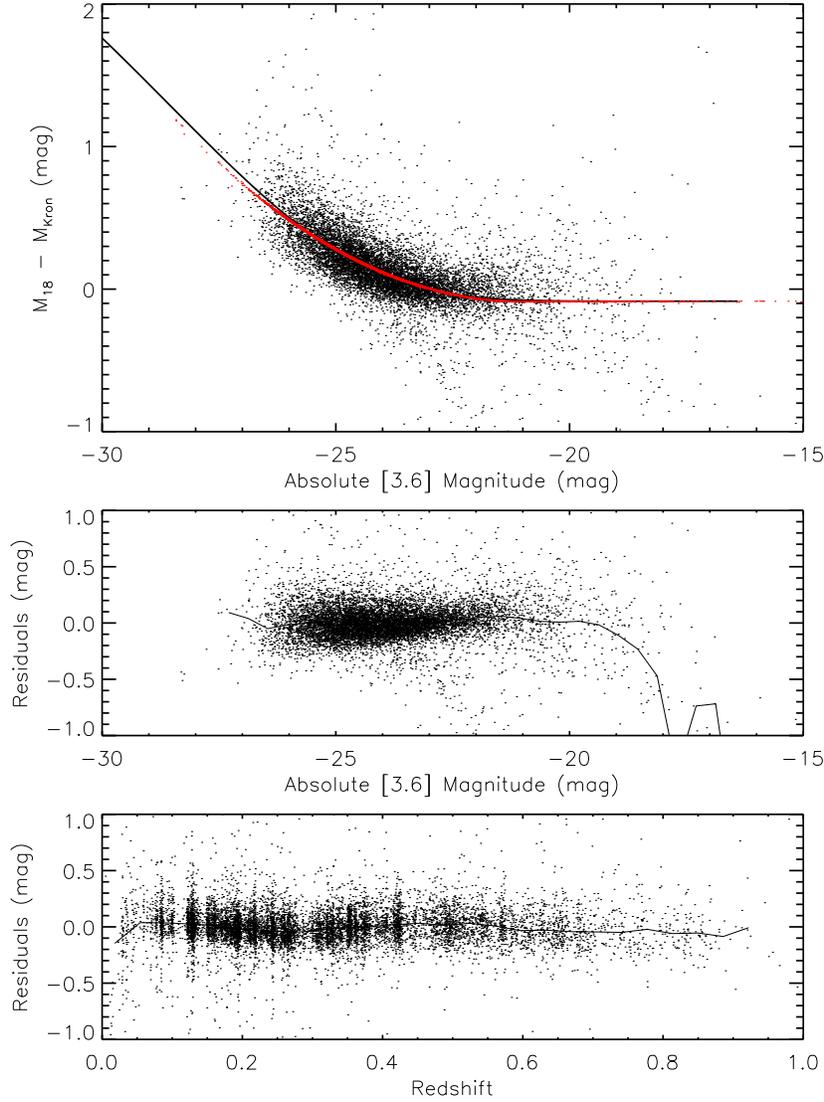}
\caption{The upper panel shows the difference between the fixed physical aperture (18\h\ kpc) magnitude and Kron magnitude, $m_{18} - m_{Kron}$, versus luminosity for the 3.6\um\ band.  The solid line shows the expected offset between Kron magnitude and $m_{18}$ assuming a de Vaucouleurs profile with $L \propto R_{eff}^{1.6}$ (Bernardi et al.\ 2003).
The black dots are the measurements, and the red dots (almost overlapping with the solid line) are the best fit model including a quadratic luminosity dependence and a linear redshift dependence.
The middle and lower panels show the residuals between the measurement and the best fit model versus luminosity and redshift, respectively, and the solid lines in the two panels show the median residuals.
There is little redshift dependence suggesting that any redshift dependent biases in the magnitudes are 
smaller than our statistical uncertainties.
\label{fig:diff}}
\end{figure}

As we attempted to measure the evolution of the LFs with redshift, it became clear that
evolving magnitude definitions could be a serious problem since a 0.1 mag drift in the 
magnitude definition from $z=0$ to 0.5 corresponds to a $\Delta Q= 0.2$ change in the 
evolution rate.  The existence of some problems was easily diagnosed by redshift dependent changes
in Kron radii between bands, and the behavior of aperture versus Kron magnitudes.  
The cleanest test for evolution is to synthesize metric aperture magnitudes subtending a fixed physical scale,
as these should have no redshift dependence if there is no evolution.  
We use this method to test whether there are additional corrections needed beyond those discussed 
in \S~\ref{sec:sample}.
Unfortunately, finite
metric apertures sample different fractions of galaxies depending on their luminosity (size),
so we must model the luminosity dependence while searching for a redshift dependence.
Figure~\ref{fig:diff} shows the difference, $m_{18} - m_{Kron}$, between the $18h^{-1}$ kpc 
diameter metric magnitude ($m_{18}$) and Kron magnitude for the 3.6\um\ band 
both for the data and the mean difference found assuming de Vaucouleurs profile galaxies with 
$L \propto R_{eff}^{1.6}$ (Bernardi et al.\ 2003).
We see that the dominant trend comes from the luminosity.
To estimate the redshift biases, we fit
\begin{equation}
m_{18} - m_{Kron} = \left\{ \begin{array}
{r@{\quad:\quad}l} 
a(M_{Kron}-M_{0})^2 + bz + c & M_{Kron} < M_{0} \\ bz + c & M_{Kron} \ge M_{0} 
\end{array} \right.
\label{eqn:aone}
\end{equation}
using a one-sided quadratic term for the luminosity trend and adding a linear (false) evolution with
redshift.
The parameters are the mean offset $c$, the amplitude of the quadratic dependence on luminosity $a$, 
the lumonisity $M_0$ above which the metric aperture underestimate the flux, and the redshift bias $b$.
As we show in Figure~\ref{fig:diff}, the one-sided quadratic term models the mean luminosity trend well.
The resulting estimates for the redshift biases are $b=-0.02\pm0.02$ and $0.08\pm0.02$ for 3.6 and 4.5\um\ bands.
These are significantly smaller than the statistical uncertainties ($\Delta Q \sim 0.5$) in the evolution rates, so we decided to apply no further corrections.
In the 5.8 and 8.0\um\ bands, our simple model failed to reproduce the measured difference between $m_{18}$ and $m_{Kron}$ possibly due to the complexity that the PAH emission from star formation may 
not follow the de Vaucouleurs profile.  Since our tests in the 3.6 and 4.5\um\ bands
show no significant redshift dependent biases, we adopted a consistent photometry scheme in 5.8 and 8.0\um\ 
bands same as that for the 3.6 and 4.5\um\ bands.

\end{document}